\documentclass[12pt]{article}%
\usepackage{heck}
\usepackage{graphicx}
\usepackage{makeidx}
\usepackage{multicol}
\usepackage{geometry}
\usepackage{amsfonts}
\usepackage{amssymb}
\usepackage{amsmath}%
\providecommand{\U}[1]{\protect\rule{.1in}{.1in}}
\begin{document}

\date{February, 2007}

\preprint{hep-th/0702077 \\ HUTP-07/A0002 }

\institution{HarvardU}{Jefferson Physical Laboratory, Harvard University, Cambridge,
MA 02138, USA}%

\institution{MIT}%
{Center for Theoretical Physics, MIT, Cambridge, MA 02139, USA}%
%

\title{Phase Structure of a Brane/Anti-Brane System at Large $N$}%
%

\authors{Jonathan J. Heckman\worksat{\HarvardU,}\footnote{e-mail: {\tt
jheckman@fas.harvard.edu}},
Jihye Seo\worksat{\HarvardU,}\footnote{e-mail: {\tt jihyeseo@fas.harvard.edu}}
and Cumrun Vafa\worksat{\HarvardU,\MIT}\footnote{e-mail: {\tt
vafa@physics.harvard.edu}}}%

\abstract{We further analyze a class of recently studied metastable string vacua obtained by wrapping
D5-branes and anti-D5-branes over rigid homologous $S^2$'s of a non-compact Calabi-Yau threefold.
The large $N$ dual description is characterized by a potential for the glueball fields which is
determined by an auxiliary matrix model.  The higher order corrections to this potential produce a suprisingly
rich phase structure.  In particular, at sufficiently large 't Hooft coupling  the metastable vacua present at weak
coupling cease to exist.  This instability can already be seen by an open string two loop contribution
to the glueball potential.  The glueball potential also lifts some of the degeneracy in the vacua characterized
by the phases of the glueball fields.  This generates an exactly computable non-vanishing axion potential at
large $N$.}%

\maketitle

\section{Introduction}

At present, few exact results are available on the phase structure of
supersymmetry breaking backgrounds in either gauge or string theories.
\ Indeed, whereas the requirement of holomorphicity determines the form of
many corrections to the vacua of a supersymmetric theory, generically no such
constraint is available when supersymmetry is broken. \ It is nevertheless
natural to consider non-supersymmetric metastable vacua of a supersymmetric
theory because the underlying supersymmetry of the theory allows more control
over the dynamics of supersymmetry breaking. \ Non-supersymmetric metastable
string constructions have been studied in
\cite{VafaLargeN,KachruPearsonVerlinde,KKLT,KachruMcGreevy,KachruFranco,ABSV,VerlindeMonopole}%
. \ Recent progress in finding non-supersymmetric metastable vacua in
supersymmetric QCD-like field theories was achieved in \cite{ISS} and
subsequent string theory realizations of this work
\cite{OoguriOokouchi,FrancoUranga,MQCDSeibergShih,TatarMeta}.

On the other hand, it is by now well-established that in certain cases the
large $N$ supersymmetric dynamics of open strings at strong 't Hooft coupling
admits a holographic dual description in terms of weakly coupled closed
strings. \ Notable examples are the AdS/CFT correspondence
\cite{juanAdS,gkPol,witHolOne} and geometric transitions
\cite{VafaLargeN,KlebanovStrassler,MaldacenaNunezYangMills}. \ In this note we
exploit the fact that large $N$ \ holography is expected to be a more general
property of many non-supersymmetric gravitational systems to analyze the phase
structure of a strongly coupled supersymmetry breaking background. \ In
particular, we further study the large $N$ dual of a configuration of branes
and anti-branes with $\mathcal{N}=0$ supersymmetry of the type recently
considered in \cite{ABSV}.

More precisely, we study type IIB\ string theory compactified on the local
Calabi-Yau threefold given by a small resolution of the hypersurface defined
by:%
\begin{equation}
y^{2}=W^{\prime}(x)^{2}+uv
\end{equation}
where $W^{\prime}(x)$ is a polynomial of degree $n$ and $x,y,u,v\in%
\mathbb{C}
$. \ Our brane configuration consists of spacetime filling D5-branes and
anti-D5-branes wrapped over homologous and minimal size rigid $S^{2}$'s of the
internal geometry. \ We denote by $\left\vert N_{i}\right\vert $ the number of
branes or anti-branes wrapped over the $i^{\mathrm{th}}$ $S^{2}$. \ Here
$N_{i}$ is understood to be a positive (negative) integer for D5-branes
(anti-D5-branes). \ In the absence of branes and anti-branes, the resulting
theory in four dimensions would have preserved $\mathcal{N}=2$ supersymmetry.
\ This system is non-supersymmetric because each type of brane preserves a
different $\mathcal{N}=1$ supersymmetry. \ Even so, this configuration is
metastable because the tension of the branes generates a potential barrier
against the expansion of the $S^{2}$'s.

\begin{figure}
[ptb]
\begin{center}
\includegraphics[
height=4.8393in,
width=3.1689in
]%
{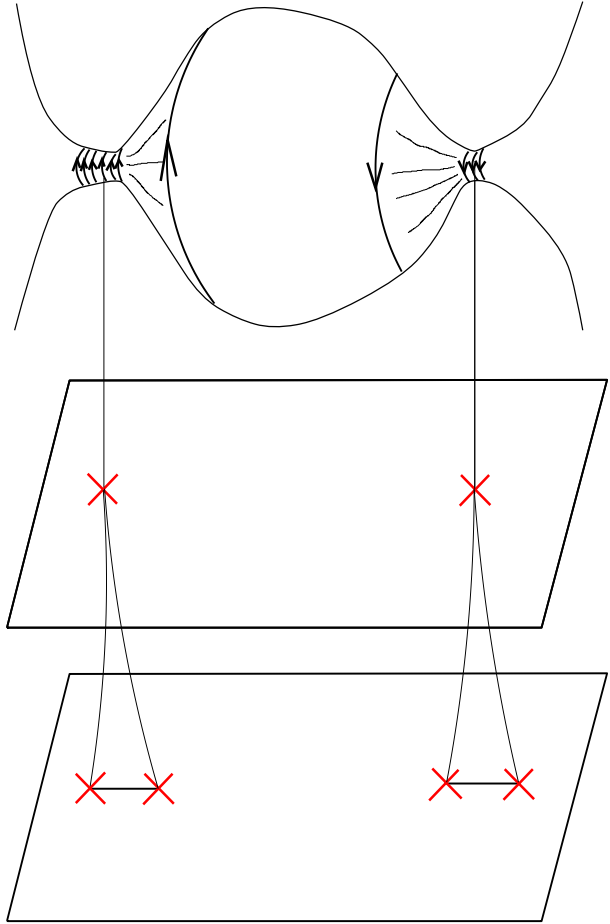}%
\caption{Depiction of the geometric transition from the open string picture
with branes wrapped over minimal size $S^{2}$'s (top) to the large $N$ closed
string dual where the branes and homologous $S^{2}$'s have been replaced by
flux threading topologically distinct $S^{3}$'s of a new geometry (bottom).
\ The lines with red crosses denote finite length branch cuts on the Riemann
surface of the local geometry after the transition. \ In the open string
picture the area of the $S^{2}$ in the middle of the bulge is approximately
$|W^{\prime}(x)|$. \ When branes are wrapped over these minimal size $S^{2}%
$'s, the bare tension of the branes creates a potential barrier against
brane/anti-brane annihilation.}%
\label{2cut3sheets}%
\end{center}
\end{figure}

In the holographic dual theory the original branes and anti-branes wrapping
$S^{2}$'s are replaced with flux threading topologically distinct $S^{3}$'s of
a new geometry described by a hypersurface of the form:%
\begin{equation}
y^{2}=W^{\prime}(x)^{2}+b_{n-1}x^{n-1}+\cdots+b_{0}+uv
\end{equation}
where the $b_{i}$ are normalizable complex deformations of the singular
geometry. \ The $x$ and $y$ coordinates define a Riemann surface fibered over
the coordinates $u$ and $v$. \ As shown in {figure (\ref{2cut3sheets})}, each
$S^{3}$ of the new Calabi-Yau reduces to a contour encircling a finite length
branch cut of the complex $x$-plane. \ The size of each $S^{3}$ is controlled
by a flux induced effective potential. \ Supersymmetric configurations of this
type have been studied in \cite{CIV}. \ It was conjectured in \cite{ABSV} that
this same geometric transition remains valid for non-supersymmetric
configurations. \ In this case, the vacuum is determined by the full potential
rather than the superpotential. \ Rather importantly, as opposed to a generic
$\mathcal{N}=1$ theory with broken supersymmetry, at leading order in
$1/N$,\ the form of the K\"{a}hler potential is fixed by the special geometry
of the manifold\footnote{As explained in \cite{ABSV}, this is due to the fact
that at leading order in $1/N$, the fluxes spontaneously break $\mathcal{N}=2$
supersymmetry.}. \ For sufficiently low flux quanta the size of each $S^{3}$
is stabilized at a small value \cite{ABSV}. \ Throughout this paper we will
refer to this location in moduli space as the \textit{semi-classical expansion
point}.%

But as we increase the 't Hooft coupling the $S^{3}$'s will expand in size so
that higher order corrections to the effective potential will play a more
important r\^{o}le in determining the vacuum of the theory. \ These
corrections are efficiently summarized by an auxiliary matrix model
\cite{DijkgraafVafaI,DijkgraafVafaII,DijkgraafVafaIII} which determines the
classical periods of the closed string dual. \ One of the purposes of this
paper is to show that such corrections generate an intricate phase structure
which is absent in the supersymmetric case.

When supersymmetry is broken, a metastable system of fluxes will eventually
lower its energy by annihilating flux lines. \ Rather than treating an
individual system, it is therefore more appropriate to treat the totality of
all possible flux configurations which admit metastable vacua. \ To this end,
we invert the question of finding critical points of $V_{\mathrm{eff}}$ and
instead ask: Given a point in moduli space, which brane/anti-brane
configurations would stabilize the moduli at this geometry? \ Framing the
question this way, we find necessary conditions for the existence of mutually
non-supersymmetric metastable configurations of D5- and anti-D5-branes wrapped
over the $S^{2}$'s of the original geometry. \ We next analyze the phase
structure of the two cut geometry.

Except when explicitly noted, the remainder of our results apply to the two
cut geometry with the corresponding $S^{3}$'s supported by purely RR three
form flux. \ By determining the vacua of some special flux configurations, we
argue that $N_{1}$, $N_{2}$ and $\theta_{\mathrm{YM}}$ respectively control
the sizes and relative orientation of the cuts. \ For $N_{1}=-N_{2}$ and
$\theta_{\mathrm{YM}}$ arbitrary, we find metastable vacua such that the
branch cuts are mirror reflections across an axis of the complex $x$-plane.
$\ $For $N_{1}$ and $N_{2}$ arbitrary and $\theta_{\mathrm{YM}}=0$, we find
metastable vacua such that the branch cuts align along a common axis of the
complex $x$-plane.

Even so, both supersymmetric and non-supersymmetric flux configurations admit
many other potentially metastable vacua. \ Indeed, even non-supersymmetric
$U(N)$ Yang-Mills theory admits $O(N)$ metastable vacua which become exactly
stable in the $N=\infty$ limit \cite{WittenLargeNChirDyn,WittenthetaSolution}.
\ Heuristically, these vacua are non-supersymmetric analogues of the
$\mathrm{Tr}(-1)^{F}=N$ energetically degenerate supersymmetric confining
vacua present in pure $\mathcal{N}=1$ $SU(N)$ super Yang-Mills theory
\cite{ShifmanTunneling}. \ For multi-cut geometries these other vacua are
obtained at leading order in the closed string dual by rotating the branch
cuts by a discrete angle $\pi/\left\vert N_{i}\right\vert $. \ Although this
geometric symmetry is deformed by higher order corrections, the corresponding
supersymmetric confining vacua remain exactly degenerate in energy.

By contrast, we find that for non-supersymmetric confining vacua, the two loop
contribution to the potential lifts this degeneracy so that it is
energetically favorable for the branch cuts to align along a common axis.
\ This corresponds to an alignment of phases in the glueball fields.
\ Physically this follows from the fact that the potential energy due to the
Coulomb attraction between the branes and anti-branes is lowest for such a
configuration. \ We find that the energy dependence of the vacuum as a
function of $\theta_{\mathrm{YM}}$ agrees with the form expected in large $N$
gauge theories\footnote{Upon imbedding our non-compact geometry into a compact
Calabi-Yau threefold, this realignment generates a potential for the axion.
\ Although we do not develop this into a fully viable phenomenological model,
we believe that this mechanism may be of independent interest for solving the
strong CP\ problem.}. \ In addition, we also find a large number of
potentially metastable vacua in accord with expectations from large $N$
arguments. \ Indeed, although there is a single energetically preferred
confining vacuum, the rate of decay to this lowest energy configuration is
suppressed by large $N$ effects. \ Thus, once we have established the
existence of a \textit{single} metastable brane/anti-brane configuration,
general arguments from large $N$ gauge theories imply the presence of a large
number of \textit{additional} potentially metastable vacua.

Restricting further to the cases $N_{1}=-N_{2}$ and $\left\vert N_{1}%
\right\vert \gg\left\vert N_{2}\right\vert $ with the branch cuts aligned with
the energetically preferred configuration along the real axis of the complex
$x$-plane, we find that for sufficiently large 't Hooft coupling, but far
before the cuts touch, the theory undergoes a phase transition which lifts the
metastable vacua present at weak coupling. \ This is a novel phenomenon where
strong coupling effects lead to a loss of stability in a classically
metastable brane/anti-brane system.

Once the 't Hooft coupling is large and metastability is lost, we can ask
about the fate of the vacuum. \ In order to address this, it is necessary to
go beyond the regime where a perturbative computation of the potential is
valid. Since we have the exact potential at large $N$, we can study this
regime as well. \ Using a combination of numerical and analytic arguments, we
find that the dynamics of the fluxes drive the moduli to a configuration where
the branch cuts nearly touch. \ Close to this region in moduli space, a gas of
nearly tensionless domain walls will typically cause the system to tunnel to a
metastable vacuum of lower flux. \ When this does not occur, the cuts can
touch and additional light magnetic states condense. \ In this case, we find
that the resulting geometry is a non-K\"{a}hler manifold.

The organization of the rest of this paper is as follows. \ In section
\ref{review} we establish notation and review the conjecture of \cite{ABSV} on
the large $N$ dual of spacetime filling D5-branes and anti-D5-branes wrapped
over $S^{2}$'s of a non-compact Calabi-Yau threefold. \ In section
\ref{critpoints} we derive necessary conditions for the existence of a
metastable vacuum. \ Beginning in section \ref{twocut} we specialize to the
two cut geometry and explain how the fluxes control the sizes and relative
orientation of the branch cuts. \ We next show in section \ref{confiningvacua}
that a two loop effect lifts the degeneracy in energy between the confining
vacua of the theory so that it is energetically favorable for the branch cuts
(i.e. the phases of the glueball condensates) to align along a common axis.
\ In this same section we also find a large number of additional potentially
metastable vacua and compare the $\theta_{YM}$ dependence of the vacuum energy
density with general expectations from large $N$ gauge theories. \ In section
\ref{Breakdown} we show that for sufficiently large 't Hooft coupling, this
two loop effect lifts the metastable vacua present near the semi-classical
expansion point. \ This causes the branch cuts to expand until they nearly
touch. \ Section \ref{Endpoint} discusses the behavior of the system near this
region of moduli space, and section \ref{Conclusions} presents our conclusions
and possible avenues of further investigation.

\section{Geometrically Induced Metastability\label{review}}

In this section we set our notation and discuss in more detail the large $N$
dual description of the metastable brane/anti-brane configuration we shall
study in this paper. \ The open string description of our system consists of
D5-branes and anti-D5-branes which fill Minkowski space and wrap $n$ minimal
size $S^{2}$'s of a local Calabi-Yau threefold defined by the hypersurface:%
\begin{equation}
y^{2}=W^{\prime}(x)^{2}+uv \label{rescon}%
\end{equation}
where $x,y,u,v\in%
\mathbb{C}
$ and $W^{\prime}(x)\equiv g(x-a_{1})\cdots(x-a_{n})$ is a degree $n$
polynomial. \ Because the geometry is non-compact, the $a_{i}$ correspond to
non-normalizable modes which determine the relative separation between the
branes. \ The minimal size $S^{2}$'s of the geometry are all homologous and
are located at the points where $W^{\prime}(x)$ vanishes. \ Indeed, at a
generic point of the complex $x$-plane the area of an $S^{2}$ is given by the
relation:%
\begin{equation}
A(x)=\left(  \left\vert W^{\prime}(x)\right\vert ^{2}+\left\vert r\right\vert
^{2}\right)  ^{1/2}%
\end{equation}
where $r$ denotes the size of the $S^{2}$ at $x=a_{i}$. \ Because the branes
and anti-branes preserve different supersymmetries, the corresponding system
does not preserve any supersymmetry. \ While there is no topological
obstruction to the branes and anti-branes annihilating, the system is
nevertheless \textit{geometrically} metastable because the bare tension of the
branes produces a potential barrier against the expansion of the branes. \ See
{figure (\ref{2cut3sheets})} for the local behavior of this configuration.

\begin{figure}
[ptb]
\begin{center}
\includegraphics[
height=2.0481in,
width=4.4342in
]%
{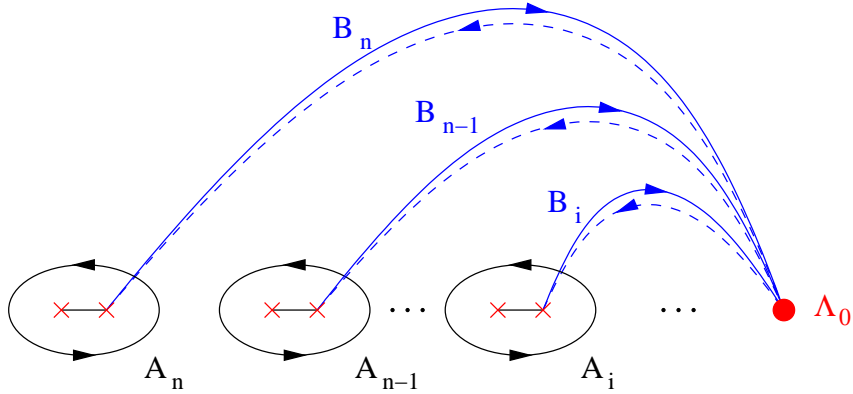}%
\caption{Depiction of the complex $x$-plane corresponding to the Riemann
surface defined by equation (\ref{defcon}) with $uv=0$. \ The compact
$A$-cycles reduce to counterclockwise contours which encircle each of the $n$
branch cuts of the Riemann surface. \ The non-compact $B$-cycles reduce to
contours which extend from $x=\Lambda_{0}$ on the lower sheet (dashed lines)
to $x=\Lambda_{0}$ on the upper sheet\ (solid lines).}%
\label{Riemann}%
\end{center}
\end{figure}

In the holographic dual description, the branes and anti-branes wrapping $n$
homologous $S^{2}$'s of the original geometry are replaced by fluxes threading
the $n$ topologically distinct $S^{3}$'s of the new geometry. \ The local
Calabi-Yau threefold after the transition is defined by the equation:%
\begin{equation}
y^{2}=W^{\prime}(x)^{2}+b_{n-1}x^{n-1}+\cdots+b_{0}+uv \label{defcon}%
\end{equation}
where the $b_{i}$ correspond to the $n$ normalizable complex deformation
parameters of the Calabi-Yau. \ This complex equation defines a two-sheeted
Riemann surface fibered over the $u$ and $v$ coordinates. \ The $b_{i}$ split
the double roots of $W^{\prime}(x)^{2}$, creating $n$ finite length branch
cuts on the complex $x$-plane of the Riemann surface. \ See
figure\ (\ref{Riemann}) for a depiction of this geometry.

The $n$ $S^{3}$'s correspond to $n$ 3-cycles $A_{i}$ such that $A_{i}\cap
A_{j}=0$ for all $i,j$. \ Dual to each $A$-cycle is a non-compact $B$-cycle
such that $A_{i}\cap B_{j}=-B_{j}\cap A_{i}=\delta_{ij}$ and $B_{i}\cap
B_{j}=0$ for all $i,j$. \ At the level of the Riemann surface, the $A_{i}$
reduce to $n$ distinct counter-clockwise contours encircling each of the $n$
branch cuts of the Riemann surface and the $B_{i}$ reduce to contours which
extend from the point $x=\Lambda_{0}$ on the lower sheet to the point
$x=\Lambda_{0}$ on the upper sheet. \ The IR\ cutoff defined by $\Lambda_{0}$
in the geometry is identified with a UV\ cutoff in the open string
description. \ The periods of the holomorphic three form $\Omega$ along the
cycles $A_{i}$ and $B_{i}$ define a basis of special coordinates for the
complex structure moduli space:%
\begin{equation}
S_{i}=\underset{A_{i}}{\int}\Omega,\text{ \ \ \ \ }\Pi_{i}=\text{\ }%
\frac{\partial\mathcal{F}_{0}}{\partial S_{i}}=\underset{B_{i}}{\int}\Omega
\end{equation}
where $\mathcal{F}_{0}$ denotes the genus zero prepotential. \ In the absence
of fluxes, each $S_{i}$ corresponds to the scalar component of a $U(1)$
$\mathcal{N}=2$ vector multiplet. \ Once branes are introduced, each $S_{i}$
is identified in the open string description with the size of a gaugino
condensate. \ Defining the period matrix:%
\begin{equation}
\tau_{ij}=\frac{\partial\Pi_{i}}{\partial S_{j}}=\frac{\partial^{2}%
\mathcal{F}_{0}}{\partial S_{i}\partial S_{j}},
\end{equation}
to leading order in the $1/N$ expansion, the K\"{a}hler metric for the
effective field theory is $\operatorname{Im}\tau_{ij}$. \ For future use we
also introduce the Yukawa couplings:%
\begin{equation}
\mathcal{F}_{ijk}\equiv\frac{\partial^{3}\mathcal{F}_{0}}{\partial
S_{i}\partial S_{j}\partial S_{k}}.
\end{equation}

We now describe the large $N$ dual description of brane configurations which
preserve $\mathcal{N}=1$ supersymmetry. \ To this end, recall that for
supersymmetric flux configurations the flux-induced superpotential is
\cite{GukovVafaWitten}:%
\begin{equation}
\mathcal{W}_{\mathrm{eff}}=\int H_{3}\wedge\Omega=\alpha\left(  S_{1}%
+\cdots+S_{n}\right)  +(N_{1}\Pi_{1}+\cdots+N_{n}\Pi_{n}) \label{fluxpot}%
\end{equation}
\newline where $H_{3}$ denotes the net three form flux after the system
undergoes a geometric transition:%
\begin{equation}
H_{3}=H_{\mathrm{RR}}+\tau_{\mathrm{IIB}}H_{\mathrm{NS}}\text{.}%
\end{equation}
In the above equation, $H_{\mathrm{RR}}$ is the net RR\ three form field
strength, $H_{\mathrm{NS}}$ is the net NS three form field strength and
$\tau_{\mathrm{IIB}}=C_{0}+ie^{-\phi}$ is the type IIB\ axio-dilaton.
\ Explicitly,
\begin{equation}
N_{i}=\underset{A_{i}}{\int}H_{3},\text{ \ \ \ \ \ }\alpha=\alpha
_{i}=-\underset{B_{i}}{\int}H_{3} \label{fluxes}%
\end{equation}
for all $i$. \ Note in particular that when $N_{i}=p_{i}+\tau_{\mathrm{IIB}%
}q_{i}$ is a general complex number, this describes the geometric transition
of a $(p_{i},q_{i})$ 5-brane wrapped over the $i^{\mathrm{th}}$ $S^{2}$ with
$p_{i}$ units of D5-brane charge and $q_{i}$ units of NS5-brane charge.\ In
the open string theory, the parameter $\alpha$ corresponds to minus the
complexified gauge coupling evaluated at the UV cutoff:%
\begin{equation}
\alpha=\alpha\left(  \Lambda_{0}\right)  =-\frac{\theta_{\mathrm{YM}}}{2\pi
}-\frac{4\pi i}{g_{\mathrm{YM}}^{2}}.
\end{equation}
The purely $S_{i}$ sector of the theory is described by the Lagrangian
density:%
\begin{equation}
\mathcal{L}_{S}=\frac{1}{g_{s}^{2}}\Lambda_{\mathrm{UV}}^{-4}\left(
\operatorname{Im}\tau\right)  _{i\overline{j}}\partial_{\mu}S^{i}\partial
^{\mu}\overline{S^{j}}+\Lambda_{\mathrm{UV}}^{4}V_{\mathrm{eff}}\left(
S_{i},\overline{S_{j}}\right)  \label{SLAGRANGIAN}%
\end{equation}
where $\Lambda_{\mathrm{UV}}$ is a UV\ mass scale which is potentially
different from $\Lambda_{0}$, and $V_{\mathrm{eff}}$ is given by\footnote{In
string frame, the overall $g_{s}$ scaling of the superpotential and K\"{a}hler
potential is $\mathcal{W}_{\mathrm{eff}}/g_{s}$ and $K/g_{s}^{2}$,
respectively. \ When it will not cause confusion, we shall suppress the
$g_{s}$ scaling of these two quantities in our computations.}:%
\begin{equation}
V_{\mathrm{eff}}=\partial_{k}\mathcal{W}_{\mathrm{eff}}\left(  \frac
{1}{\operatorname{Im}\tau}\right)  ^{kl}\overline{\partial_{l}\mathcal{W}%
_{\mathrm{eff}}}=\left(  \alpha_{k}+N^{k^{\prime}}\tau_{k^{\prime}k}\right)
\left(  \frac{1}{\operatorname{Im}\tau}\right)  ^{kl}\left(  \overline{\alpha
}_{l}+\overline{\tau}_{ll^{\prime}}\overline{N}^{l^{\prime}}\right)  \text{.}
\label{Veffdef}%
\end{equation}
When it will not cause any confusion, we will work in units where
$\Lambda_{\mathrm{UV}}$ is normalized to unity.

Having reviewed the large $N$ dual description for $\mathcal{N}=1$ brane
configurations, we now describe the $\mathcal{N}=0$ analogue of this
description when some of the branes are replaced by anti-branes. \ In
\cite{ABSV} it was conjectured that the form of the flux-induced effective
potential for the $S_{i}$'s is essentially unchanged from the supersymmetric
case. \ To properly compare the energy of both branes and anti-branes
simultaneously, it is appropriate to shift $V_{\mathrm{eff}}$ by a multiple of
the bare tensions of the branes \cite{ABSV}:%
\begin{equation}
V_{\mathrm{eff}}\mapsto V_{\mathrm{eff}}+\frac{8\pi}{g_{\mathrm{YM}}^{2}%
}\left(  N_{1}+\cdots+N_{n}\right)  \text{.} \label{shiftedpotential}%
\end{equation}

\subsection{Matrix Models and $V_{\mathrm{eff}}$}

In this subsection we review the connection between matrix models and special
geometry. \ As originally proposed in \cite{DijkgraafVafaI}, the genus zero
prepotential of the geometry defined by equation (\ref{defcon}) is exactly
computed by the planar limit of a large $M$ auxiliary matrix model with
partition function:%
\begin{equation}
Z_{\mathrm{MM}}=\frac{1}{\mathrm{Vol}\left(  U\left(  M\right)  \right)  }\int
d\Phi\exp\left(  -\frac{1}{g_{s}}\text{\textrm{Tr}}W\left(  \Phi\right)
\right)
\end{equation}
where $\Phi$ is a holomorphic $M\times M$ matrix and the above matrix integral
should be understood as a contour integral. \ The prepotential of the $n$-cut
geometry near the semi-classical expansion point is given by expanding the
eigenvalues of $\Phi$ about the $n$ critical points of the polynomial $W$.
\ The usual eigenvalue repulsion term of the matrix model causes these
eigenvalues to fill the $n$ cuts of the geometry after the geometric
transition. \ With $M_{i}$ eigenvalues sitting at the $i^{\mathrm{th}}$ cut of
the geometry, this matrix model may be recast as an $n$-matrix model of the
form:%
\begin{equation}
Z_{\mathrm{MM}}=\frac{1}{\underset{i=1}{\overset{n}{%
{\displaystyle\prod}
}}\mathrm{Vol}\left(  U\left(  M_{i}\right)  \right)  }\int d\Phi_{1}\cdots
d\Phi_{n}\exp\left(  -\frac{1}{g_{s}}\underset{i=1}{\overset{n}{\sum}%
}{\mathrm{Tr}}W_{i}\left(  \Phi_{i}\right)  -\frac{1}{g_{s}}{\mathrm{Tr}%
}W_{\mathrm{int}}\left(  \Phi_{1},\cdots,\Phi_{n}\right)  \right)
\end{equation}
in the obvious notation. \ We caution the reader that the numbers $M_{i}$ are
unrelated to the wrapping numbers of the branes.

The connection between the above matrix model and the special geometry of the
Calabi-Yau threefold defined by equation (\ref{defcon}) is obtained as
follows. \ The periods of the $A$-cycles are given by the partial 't Hooft
couplings of the matrix model \cite{DijkgraafVafaI}:%
\begin{equation}
S_{i}=g_{s}M_{i}\text{.}%
\end{equation}
Evaluating $Z_{\mathrm{MM}}$ in the saddle point approximation, the planar
limit of the free energy for the matrix model is identified with the genus
zero prepotential for the complex structure moduli space of the Calabi-Yau
threefold:%
\begin{equation}
\mathcal{F}_{0}=\mathcal{F}_{\mathrm{measure}}+\mathcal{F}_{\mathrm{pert}}%
\end{equation}
where $\mathcal{F}_{\mathrm{measure}}$ corresponds to contributions from the
$\mathrm{Vol}\left(  U\left(  M_{i}\right)  \right)  $ factors in the
path-integral measure \cite{OoguriVafaWorldsheet} and $\mathcal{F}%
_{\mathrm{pert}}$ corresponds to perturbative contributions from planar
Feynman diagrams:%
\begin{align}
2\pi i\mathcal{F}_{\mathrm{measure}} &  =\underset{i=1}{\overset{n}{\sum}%
}\frac{1}{2}S_{i}^{2}\log\frac{S_{i}}{{\Lambda_{0}^{3}}}\\
2\pi i\mathcal{F}_{\mathrm{pert}} &  =-\underset{i=1}{\overset{n}{\sum}}%
S_{i}W(a_{i})+\underset{0\leq i_{1},\cdots,i_{n}}{\sum}C_{i_{1}\cdots i_{n}%
}S_{1}^{i_{1}}\cdots S_{n}^{i_{n}}\text{.}%
\end{align}

From the perspective of the open string theory, the contribution
$\mathcal{F}_{\mathrm{measure}}$ reproduces the expected
Veneziano-Yankielowicz terms in the superpotential. \ At leading order in the
expansion of the periods about small $S_{i}$, these contributions serve to
stabilize the magnitude of the glueball fields at the exponentially small
value $\sim\exp\left(  -8\pi^{2}/g_{\mathrm{YM}}^{2}\left\vert N_{i}%
\right\vert \right)  $. \ In addition to this leading order behavior, the
power series in the $S_{i}$'s given by $\mathcal{F}_{\mathrm{pert}}$ produces
subleading corrections to the form of the glueball potential. \ Although the
form of such corrections are difficult to calculate for a general confining
gauge theory, in the present case the integrable structure of the matrix model
ensures that the $C_{i_{1}...i_{n}}$ are \textit{in principle calculable}.

\subsection{Leading Order Behavior\label{leadingorderbeh}}

\begin{figure}
[ptb]
\begin{center}
\includegraphics[
height=0.9805in,
width=5.7634in
]%
{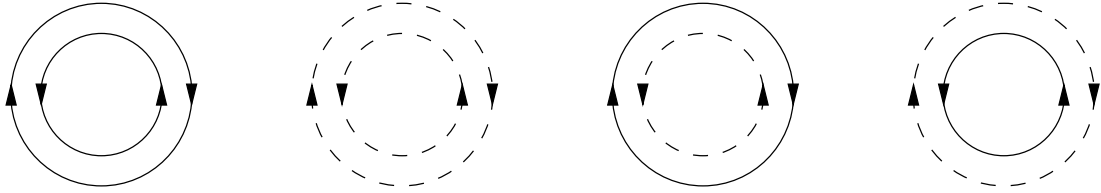}%
\caption{Topologically distinct one loop planar diagram contributions to the
prepotential for the two cut matrix model.\ \ Solid lines denote branes and
dashed lines denote anti-branes.}%
\label{oneloopprepot}%
\end{center}
\end{figure}

\begin{figure}
[ptb]
\begin{center}
\includegraphics[
height=2.479in,
width=4.005in
]%
{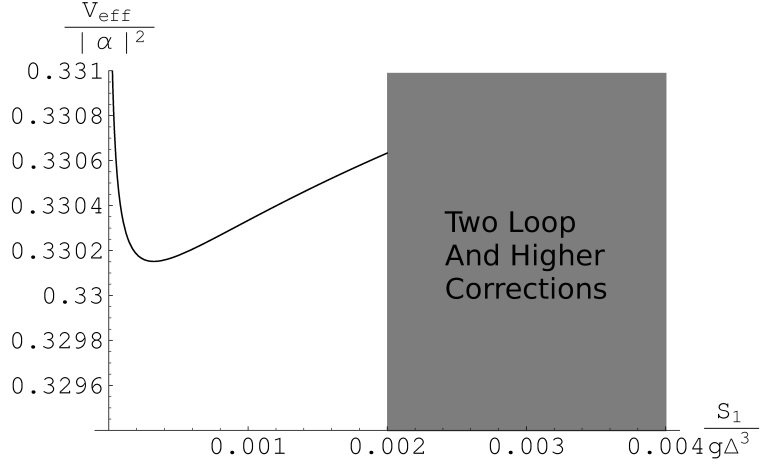}%
\caption{Plot of $V_{\mathrm{eff}}/\left\vert \alpha\right\vert ^{2}$ in the
one loop approximation along the locus $S_{1}/g\Delta^{3}=-S_{2}/g\Delta
^{3}>0$ for a flux configuration with $N_{1}=-N_{2}$. \ In this plot
$\Lambda_{0}/\Delta\sim10^{4}$ and $\left\vert N_{1}/\alpha\right\vert
\sim0.1$. \ In the neighborhood of the semi-classical expansion point there is
a single critical point which is metastable.}%
\label{ONELOOPPLOT}%
\end{center}
\end{figure}

Following \cite{ABSV}, we now review the leading order behavior of metastable
critical points of $V_{\mathrm{eff}}$ near the semi-classical expansion point.
\ Expanding the genus zero prepotential to quadratic order in the $S_{i}$'s,
the period matrix is:%
\begin{equation}
\tau_{ii}=\frac{1}{2\pi i}\log\frac{S_{i}}{W^{\prime\prime}(a_{i})\Lambda
_{0}^{2}}\text{,\ \ \ \ \ \ }\tau_{ij}=\frac{-1}{2\pi i}\log\frac{\Lambda
_{0}^{2}}{\Delta_{ij}^{2}} \label{leadingordertaus}%
\end{equation}
for all $i\neq j$. \ In the above expression $\Delta_{ij}=a_{i}-a_{j}$ is the
relative separation between the minimal cycles over which branes or
anti-branes wrap. \ From the perspective of the auxiliary matrix model, this
leading order behavior corresponds to the sum of the measure factor terms and
all one loop planar diagrams. \ These latter perturbative contributions give
rise to terms in the prepotential proportional to $S_{i}S_{j}$.\ \ See {figure
(\ref{oneloopprepot})} for the one loop contributions to the prepotential of
the two cut geometry. \
For sufficiently small $\left\vert N_{i}\right\vert $ the complex structure
moduli are stabilized at exponentially small values \cite{ABSV}:%
\begin{align}
S_{N_{i}>0}  &  =\zeta_{i}W^{\prime\prime}\left(  a_{i}\right)  \Lambda
_{0}^{2}\underset{j\neq i}{\overset{N_{j}>0}{%
{\displaystyle\prod}
}}\left(  \frac{\Lambda_{0}}{\Delta_{ij}}\right)  ^{2\left\vert \frac{N_{j}%
}{N_{i}}\right\vert }\underset{k\neq i}{\overset{N_{k}<0}{%
{\displaystyle\prod}
}}\left(  \frac{\overline{\Lambda_{0}}}{\overline{\Delta_{ij}}}\right)
^{2\left\vert \frac{N_{k}}{N_{i}}\right\vert }\exp\left(  -\frac{2\pi i\alpha
}{\left\vert N_{i}\right\vert }\right) \\
S_{N_{i}<0}  &  =\zeta_{i}W^{\prime\prime}\left(  a_{i}\right)  \Lambda
_{0}^{2}\underset{j\neq i}{\overset{N_{j}>0}{%
{\displaystyle\prod}
}}\left(  \frac{\overline{\Lambda_{0}}}{\overline{\Delta_{ij}}}\right)
^{2\left\vert \frac{N_{j}}{N_{i}}\right\vert }\underset{k\neq i}%
{\overset{N_{k}<0}{%
{\displaystyle\prod}
}}\left(  \frac{\Lambda_{0}}{\Delta_{ij}}\right)  ^{2\left\vert \frac{N_{k}%
}{N_{i}}\right\vert }\exp\left(  \frac{2\pi i\overline{\alpha}}{\left\vert
N_{i}\right\vert }\right)
\end{align}
where $\zeta_{i}$ denotes an $N_{i}^{\mathrm{th}}$ root of unity. \ As
expected, there is a mass splitting at leading order between the bosons and
fermions which explicitly demonstrates that supersymmetry is broken.
\ Finally, at leading order there is only a single critical point of the
physical potential near the semi-classical expansion point corresponding to
the metastable minimum. \ See {figure (\ref{ONELOOPPLOT}) for an example of
this behavior in the two cut geometry}.%

\subsection{Two Loop Corrections}

\begin{figure}
[ptb]
\begin{center}
\includegraphics[
height=1.7534in,
width=4.3943in
]%
{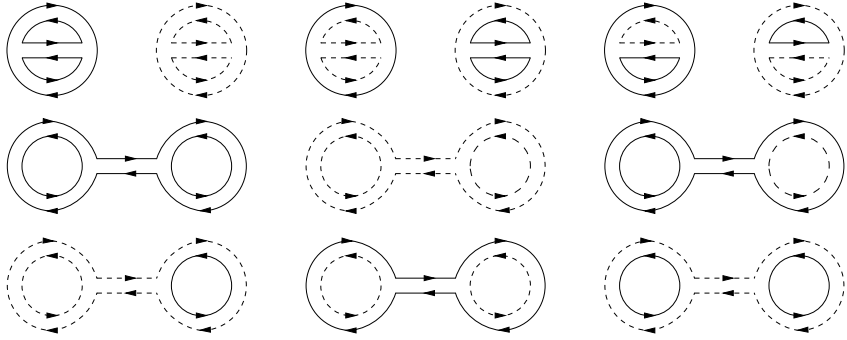}%
\caption{Two loop planar diagram contributions to the genus zero prepotential
of the two cut geometry. \ Solid lines denote branes and dashed lines denotes
anti-branes.}%
\label{twoloop}%
\end{center}
\end{figure}

We now describe two loop corrections to $V_{\mathrm{eff}}$. \ {Figure
(\ref{twoloop})} depicts the collection of topologically distinct diagrams
which contribute to the cubic term of the genus zero prepotential. \ Rather
than describe the relative contribution of each of the twelve topologically
distinct two loop planar diagrams, we merely give an example of the relevant
combinatorics. \ The combinatorial factors for the two diagrams with purely
solid lines corresponding to the disk with two holes and two disks attached by
a tube in {figure (\ref{twoloop})} are respectively $1/6$ and $1/2$. \ The
remaining two loop contributions generate additional terms at cubic order in
the genus zero prepotential and are computed in
\cite{PerturbativeMatrixModels}.%

\section{Critical Points of $V_{\mathrm{eff}}$\label{critpoints}}

In this section we derive necessary conditions that a flux configuration must
satisfy in order for the flux induced effective potential:%
\begin{equation}
V_{\mathrm{eff}}=\left(  \alpha_{k}+N^{k^{\prime}}\tau_{k^{\prime}k}\right)
\left(  \frac{1}{\operatorname{Im}\tau}\right)  ^{kl}\left(  \overline{\alpha
}_{l}+\overline{\tau}_{ll^{\prime}}\overline{N}^{l^{\prime}}\right)
\label{Veffpot}%
\end{equation}
to possess a critical point. \ Because such configurations are not absolutely
stable, the amount of flux through each $A$-cycle will decrease over a
sufficiently long time scale. \ Rather than treating one particular metastable
flux configuration, it is therefore more appropriate to treat the totality of
all flux configurations which admit metastable vacua. \ Indeed, we ask the
inverse question: Given a point in moduli space, what configuration of fluxes
would stabilize the moduli at this point? \ To this end, we solve for the
fluxes as a function of the critical point.

In what follows, we shall not require that the flux vector $N$ be an element
of an integral lattice. \ Indeed, because the critical points of the effective
potential are invariant under an overall rescaling of $\alpha$ and the $N_{i}%
$, we may approximate the $N_{i}$ as continuous parameters once they have been
rescaled to a sufficiently large value. \ Further, we shall at first allow
configurations with $N_{i}\in%
\mathbb{C}
$.

The critical points are given by differentiating the effective potential with
respect to the $S_{j}$ for $j=1,\cdots,n$:%
\begin{align}
0  &  =-2i\partial_{j}V_{\mathrm{eff}}=\left(  \alpha+N\overline{\tau}\right)
^{t}\frac{1}{\operatorname{Im}\tau}\left(  \partial_{j}\tau\right)  \frac
{1}{\operatorname{Im}\tau}\left(  \overline{\alpha}+\overline{\tau}%
\overline{N}\right) \\
&  \equiv W^{t}\left(  \partial_{j}\tau\right)  V \label{critcondition}%
\end{align}
where we have introduced two n-component column vectors $V$ and $W$:%
\begin{align}
V  &  =\frac{1}{\operatorname{Im}\tau}\left(  \overline{\alpha}+\overline
{\tau}\overline{N}\right) \\
W  &  =\frac{1}{\operatorname{Im}\tau}\left(  \alpha+\overline{\tau}N\right)
\end{align}
and for compactness we have suppressed all indices determined by matrix
multiplication. \ Solving for the n-component vectors $N$ and $\alpha$ in
terms of $V$, $W$ and $\tau$ yields:%
\begin{align}
2iN  &  =\overline{V}-W\label{Nequation}\\
2i\alpha &  =\tau W-\overline{\tau}\overline{V}. \label{alphaequation}%
\end{align}
It is therefore sufficient to express $V$ and $W$ as functions of the moduli.

Before proceeding to the solution of these equations, we now argue that for a
given point in moduli space there are at most $2^{n}$ flux configurations such
that this point is a critical point of $V_{\mathrm{eff}}$. \ In addition to
the $n$ conditions of equation (\ref{critcondition}), we obtain $n-1$
conditions from equation (\ref{alphaequation}). \ Indeed, because the column
vector $\alpha$ is proportional to the vector with 1's for all entries, the
$n-1$ dimensional subspace orthogonal to $\alpha$ is independent of both the
moduli and the fluxes. \ Letting $T_{0},\cdots,T_{n-2}$ denote the $n-1$
independent vectors which span this subspace, the dot product of the $T_{i}$
with equation (\ref{alphaequation}) yields an additional $n-1$ equations:%
\begin{equation}
0=W^{t}\tau T_{i}-\overline{V}^{t}\overline{\tau}T_{i} \label{extras}%
\end{equation}
for $i=0,\cdots,n-2$.

Combining equations (\ref{critcondition}) and (\ref{extras}) yields a total of
$2n-1$ complex equations for $2n$ complex variables. \ Because the critical
points are insensitive to the rescaling of the $n+1$ component row vector
$\left(  \alpha,N_{1},\cdots,N_{n}\right)  $, we conclude that up to an
overall rescaling by a complex number, there are a finite number of fluxes
which satisfy the required conditions. \ Finally, because the system consists
of $n-1$ linear equations and $n$ quadratic equations, the number of
\textquotedblleft critical fluxes\textquotedblright\ at a given point in
moduli space is at most $2^{n}$.

\subsection{Non-Supersymmetric Solutions\label{NONSUSYSOL}}

We now restrict our attention to flux configurations which break
supersymmetry. \ In this case, the vectors $V$ and $W$ each have at least one
non-zero component. \ To isolate the projective nature of the solutions, we
introduce affine versions of $V$ and $W$ which are completely fixed by the
moduli dependent matrices $\tau$ and $\partial_{i}\tau$. \ Without loss of
generality, we may take the non-zero component of $V$ to be $V_{1}$ and that
of $W$ to be $W_{n}$. \ Now define rescaled $n$-component vectors $\nu$ and
$\omega$ such that their complex conjugates satisfy:%
\begin{align}
V  &  =V_{1}\overline{\nu}\equiv V_{1}\left\vert \overline{\nu}\right\rangle
\text{ \ \ \ \ \ \ \ \ \ \ \ \ \ \ \ \ \ }V^{t}=V_{1}\left\langle
\nu\right\vert \label{first}\\
W  &  =W_{n}\overline{\omega}\equiv W_{n}\left\vert \overline{\omega
}\right\rangle \text{ \ \ \ \ \ \ \ \ \ \ \ \ \ }W^{t}=W_{n}\left\langle
\omega\right\vert . \label{second}%
\end{align}
where to reduce notational clutter we have switched to bra and ket notation.
\ The $2n-2$ non-trivial components of $\nu$ and $\omega$ are therefore fixed
as functions of the moduli by the $2n-2$ equations:%
\begin{align}
0  &  =\left\langle \omega\right\vert \partial_{j}\tau\left\vert \overline
{\nu}\right\rangle \label{one}\\
0  &  =\left\langle \omega\right\vert \tau\left\vert T_{0}\right\rangle
\left\langle \overline{\nu}\right\vert \overline{\tau}\left\vert
T_{k}\right\rangle -\left\langle \overline{\nu}\right\vert \overline{\tau
}\left\vert T_{0}\right\rangle \left\langle \omega\right\vert \tau\left\vert
T_{k}\right\rangle \label{two}%
\end{align}
where $j=1,\cdots,n$ and $k=1,\cdots,n-2$ with $N$ and $\alpha$ given by:%
\begin{align}
\frac{2i}{C}\left\vert N\right\rangle  &  =\left\vert \nu\right\rangle
\left\langle \omega\right\vert \tau\left\vert T_{0}\right\rangle -\left\vert
\overline{\omega}\right\rangle \left\langle \overline{\nu}\right\vert
\overline{\tau}\left\vert T_{0}\right\rangle \label{firstform}\\
\frac{2i}{C}\left\vert \alpha\right\rangle  &  =\tau\left\vert \overline
{\omega}\right\rangle \left\langle \overline{\nu}\right\vert \overline{\tau
}\left\vert T_{0}\right\rangle -\overline{\tau}\left\vert \nu\right\rangle
\left\langle \omega\right\vert \tau\left\vert T_{0}\right\rangle \label{alpha}%
\end{align}
where $C$ is an non-zero complex constant which is undetermined by the
equations. \ Because the form of these equations is somewhat similar to those
obtained in the study of the attractor equations of Calabi-Yau black holes
\cite{AttractorMech,AttStrom,AttFerrKallone,AttFerrKallTwo}, we will loosely
refer to the above as our \textit{attractor-like equations}.

\subsubsection{Brane Types and the Real Flux
Locus\label{branetypesandrealflux}}

Our goal is to study the large $N$ dual description of a system of D5-branes
and anti-D5-branes. \ Note, however, that the attractor-like equations
(\ref{firstform}) and (\ref{alpha}) indicate that every point in moduli space
is a critical point of $V_{\mathrm{eff}}$ for some configuration of $N_{i}\in%
\mathbb{C}
$. \ To impose the additional requirement that the $N_{i}$ describe D5- and
anti-D5-branes, we must further require that all of the ratios $N_{i}/N_{j}$
be real numbers. \ This defines an $n+1$ real dimensional subspace inside the
$2n$ real dimensional moduli space.

Although a complete characterization of the real flux locus is non-trivial, a
partial description exists in the special case when the $n$ cuts of the
geometry are all aligned along the real axis of the complex $x$-plane (so that
the coefficients of the polynomial defining the Calabi-Yau threefold are all
real)\ and with $\Lambda_{0}$ chosen so that the $\tau_{ij}$ are all purely
imaginary. \ We now show that in this case the ratios $N_{i}/N_{j}$ are all
real and that $\theta_{\mathrm{YM}}=0$. \ First note that there exists a
finite neighborhood around the semi-classical expansion point such that
$\tau_{ij}$ and $\mathcal{F}_{ijk}$ are pure imaginary. \ Combining this with
the leading order behavior of the $S_{i}$'s described in subsection
\ref{leadingorderbeh}, it follows that there exists a neighborhood around the
semi-classical expansion point such that the operators $\left\vert
\nu\right\rangle \left\langle \omega\right\vert $ and$\ \left\vert
\overline{\omega}\right\rangle \left\langle \overline{\nu}\right\vert $
correspond to matrices with real entries. \ By inspection of equations
(\ref{firstform}) and (\ref{alpha}), this implies that $N_{i}/N_{j}$ is real
and $\alpha/N_{i}$ is imaginary for all $i,j$. \ Note that as expected, the
$N_{i}$ control the sizes of the cuts and the parameter $\theta_{\mathrm{YM}}$
controls the relative orientation of the cuts in the geometry.

\subsubsection{Example: Two Cut Geometry}

With notation as in section \ref{NONSUSYSOL}, we have:%
\begin{align}
\left\vert T_{0}\right\rangle  &  =\left[
\begin{array}
[c]{c}%
1\\
-1
\end{array}
\right]  ,\text{ }\left\vert \nu\right\rangle =\left[
\begin{array}
[c]{c}%
1\\
\overline{V}_{2}/\overline{V_{1}}%
\end{array}
\right]  ,\text{ }\left\vert \omega\right\rangle =\left[
\begin{array}
[c]{c}%
\overline{W_{1}}/\overline{W_{2}}\\
1
\end{array}
\right]  ,\\
\left\vert \nu\right\rangle \left\langle \omega\right\vert  &  =\left[
\begin{array}
[c]{cc}%
\rho_{w} & 1\\
\overline{\rho_{v}}\rho_{w} & \overline{\rho_{v}}%
\end{array}
\right]  ,\text{ }\left\vert \overline{\omega}\right\rangle \left\langle
\overline{\nu}\right\vert =\left[
\begin{array}
[c]{cc}%
\rho_{w} & \overline{\rho_{v}}\rho_{w}\\
1 & \overline{\rho_{v}}%
\end{array}
\right]
\end{align}
where we have introduced:%
\begin{align}
\rho_{v}  &  \equiv V_{2}/V_{1}=-\frac{d_{3}\pm\sqrt{d_{3}^{2}-4d_{1}d_{2}}%
}{2d_{2}}\label{nuequation}\\
\rho_{w}  &  \equiv W_{1}/W_{2}=-\frac{d_{3}\pm\sqrt{d_{3}^{2}-4d_{1}d_{2}}%
}{2d_{1}} \label{omegaequation}%
\end{align}
with:%
\begin{equation}
d_{1}=\det\left[
\begin{array}
[c]{cc}%
\mathcal{F}_{111} & \mathcal{F}_{112}\\
\mathcal{F}_{112} & \mathcal{F}_{221}%
\end{array}
\right]  ,\text{ }d_{2}=\det\left[
\begin{array}
[c]{cc}%
\mathcal{F}_{112} & \mathcal{F}_{221}\\
\mathcal{F}_{221} & \mathcal{F}_{222}%
\end{array}
\right]  ,\text{ }d_{3}=\det\left[
\begin{array}
[c]{cc}%
\mathcal{F}_{111} & \mathcal{F}_{112}\\
\mathcal{F}_{221} & \mathcal{F}_{222}%
\end{array}
\right]  . \label{dets}%
\end{equation}
The $\pm$ signs of equations (\ref{nuequation}) and (\ref{omegaequation}) are
correlated. \ The requirement that $g_{YM}^{2}>0$ leads to an unambiguous
assignment of brane type for each branch. \ Switching from the $+$ to the $-$
branch of equations (\ref{nuequation}) and (\ref{omegaequation}) changes all
branes (anti-branes) into anti-branes (branes).

\section{Fluxes and Geometry\label{twocut}}

Unless explicitly noted, in the rest of this paper we restrict our analysis to
the phase structure of the two cut geometry with the corresponding $S^{3}$'s
supported by purely RR three form flux satisfying the condition $N_{1}%
/N_{2}<0$. \ In this section we explain in more detail how $N_{1}$, $N_{2}$
and $\theta_{\mathrm{YM}}$ respectively control the sizes and relative
orientation of the branch cuts.

Recall that the two cut geometry is given by the defining equation:%
\begin{equation}
y^{2}=W^{\prime}(x)^{2}+b_{1}x+b_{0}+uv \label{riemann}%
\end{equation}
where $W^{\prime}(x)=g(x-a_{1})(x-a_{2})$ and $b_{1}$ and $b_{0}$ control the
sizes and orientations of the branch cuts by splitting the double roots
$a_{i}$ to $a_{i}^{\pm}$. \ When $uv=0$, this is the defining equation for an
elliptic curve. \ Without loss of generality, we may take the $a_{i}$ to be
real numbers such that $a_{1}>a_{2}$. \ When the cuts are small, we have
\cite{CIV}:%
\begin{align}
S_{1}  &  \simeq\frac{g}{32}\left(  a_{1}^{+}-a_{1}^{-}\right)  ^{2}\left(
a_{1}^{+}+a_{1}^{-}-a_{2}^{+}-a_{2}^{-}\right) \label{sone}\\
-S_{2}  &  \simeq\frac{g}{32}\left(  a_{2}^{+}-a_{2}^{-}\right)  ^{2}\left(
a_{1}^{+}+a_{1}^{-}-a_{2}^{+}-a_{2}^{-}\right)  \text{.} \label{stwo}%
\end{align}
It thus follows that rotating the branch cuts in the complex $x$-plane changes
the phases of the $S_{i}$'s. \ Note that equations (\ref{sone}) and
(\ref{stwo}) are corrected at higher order by a real analytic power series in
the $a_{i}^{\pm}$. \ When $S_{1}>0>S_{2}$, the branch cuts of the geometry lie
on the real axis of the $x$-plane. \ Finally, for future use we set
$\Delta=a_{1}-a_{2}$. \ 

It follows from the discussion in subsection \ref{branetypesandrealflux} that
the space of critical points which satisfy the condition $N_{1}/N_{2}<0$
defines a three real dimensional subspace of the four real dimensional
subspace locally described by the coordinates $S_{1}$ and $S_{2}$.

Although an exact characterization of this subspace is beyond our reach, we
can still provide a crude sketch by considering various special limits. \ At
leading order in the expansion of the periods, the critical points of the
effective potential for $N_{1}>0>N_{2}$ are \cite{ABSV}:%
\begin{align}
\frac{S_{1}}{g\Delta^{3}}  &  =\zeta_{1}\left(  \frac{\Lambda_{0}}{\Delta
}\right)  ^{2}\left(  \frac{\overline{\Lambda_{0}}}{\overline{\Delta}}\right)
^{2\left\vert \frac{N_{2}}{N_{1}}\right\vert }\exp\left(  \frac{2\pi i\alpha
}{\left\vert N_{1}\right\vert }\right)  \text{ \ \ \ \ \ }\label{leadstwo}\\
-\frac{S_{2}}{g\Delta^{3}}  &  =\zeta_{2}\left(  \frac{\Lambda_{0}}{\Delta
}\right)  ^{2}\left(  \frac{\overline{\Lambda_{0}}}{\overline{\Delta}}\right)
^{2\left\vert \frac{N_{1}}{N_{2}}\right\vert }\exp\left(  -\frac{2\pi
i\overline{\alpha}}{\left\vert N_{2}\right\vert }\right)  \label{leadsthree}%
\end{align}
where the $\zeta_{i}$ are $N_{i}^{\mathrm{th}}$ roots of unity for $i=1,2$ and
label the distinct confining vacua of the low energy theory. \ Observe that
$N_{1}$ and $N_{2}$ determine the magnitudes of $S_{1}$ and $S_{2}$. \ The
$N_{i}$ therefore determine the sizes of the branch cuts in the closed string
dual. \ Further, $\theta_{\mathrm{YM}}$ controls the relative phases of
$S_{1}$ and $S_{2}$. \ Indeed, as $\theta_{\mathrm{YM}}$ varies, the branch
cuts rotate in opposite directions. \ In this section we shall assume for
simplicity that $\zeta_{1}=\zeta_{2}=1$. \ This will necessarily limit the
scope of our analysis. \ We will return to this important point later on in
section \ref{confiningvacua} where we will show that there is an energetically
preferred confining vacuum corresponding to both of the branch cuts aligned
along the real axis of the complex $x$-plane.

In the next two subsections we show that the flux configuration $N_{1}=-N_{2}$
admits metastable critical points on a $%
\mathbb{Z}
_{2}$ symmetric locus in moduli space where $S_{1}=-\overline{S_{2}}$. \ In
this case the branch cuts are of equal size and are mirror reflections across
the line halfway between $a_{1}$ and $a_{2}$. \ We next show that flux
configurations given by $N_{1}$, $N_{2}$ real and $\theta_{\mathrm{YM}}=0$
admit metastable critical points with $-S_{2},S_{1},\Lambda_{0}>0$. \ In this
case the branch cuts are of different sizes but are both aligned along the
real axis of the $x$-plane.

\subsection{Geometry of the $%
\mathbb{Z}
_{2}$ Symmetric Locus\label{z2symmloc}}

We now consider the geometry of the locus $S_{1}=-\overline{S_{2}}$. \ Setting
$uv=0$, the Riemann surface defined by equation (\ref{riemann}) is invariant
under the mapping $\sigma$:%
\begin{equation}
x\mapsto-\overline{x},\text{ }y\mapsto-\overline{y}%
\end{equation}
provided that $g,b_{0},ib_{1}\in%
\mathbb{R}
$, and $a_{1}=-\overline{a_{2}}$.

We show that invariance under this $%
\mathbb{Z}
_{2}$ symmetry implies $S_{1}=-\overline{S_{2}}$. \ The $S_{i}$'s reduce to
line integrals on the Riemann surface:%
\begin{equation}
S_{1}=\frac{1}{2\pi i}\underset{A_{1}}{%
{\displaystyle\oint}
}ydx,\text{ \ \ \ \ \ }S_{2}=\frac{1}{2\pi i}\underset{A_{2}}{%
{\displaystyle\oint}
}ydx
\end{equation}
where by abuse of notation we let the $A_{i}$ also refer to the reduction of
the $A$-cycles to two counterclockwise oriented closed loops encircling the
two branch cuts of the geometry. \ Whereas the differential element $y{d}x$ is
by construction invariant under the map $\sigma$, the 1-cycles $A_{1}$ and
$A_{2}$ transform as:%
\begin{equation}
\sigma\left(  A_{1}\right)  =-A_{2},\text{ \ \ \ \ }\sigma\left(
A_{2}\right)  =-A_{1}.
\end{equation}
It therefore follows that:%
\begin{equation}
S_{1}+\overline{S_{2}}=0.
\end{equation}
On the other hand, it follows from a general residue computation that
\cite{CIV}:%
\begin{equation}
S_{1}+S_{2}=-\frac{1}{4g}b_{1}\text{.}%
\end{equation}
We therefore conclude that when $S_{1}=-S_{2}>0$, the branch cuts are of equal
size and are aligned along the real axis of the $x$-plane. \ More generally,
$b_{0}$ (resp. $b_{1}$) predominantly controls the size (resp. relative
orientation) of the branch cuts.%
\begin{figure}
[ptb]
\begin{center}
\includegraphics[
height=1.2561in,
width=4.9041in
]%
{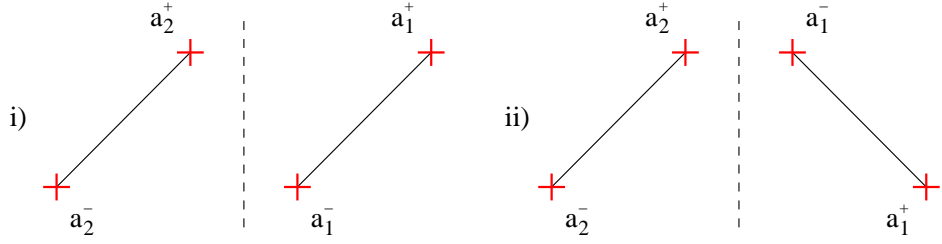}%
\caption{Depiction of the relative orientations of the branch cuts in the two
cut geometry for: i) $S_{1}=-S_{2}$ and ii) $S_{1}=-\overline{S_{2}}$.}%
\label{susyantissycuts}%
\end{center}
\end{figure}

\subsection{Flux Configurations of the $%
\mathbb{Z}
_{2}$ Symmetric Locus\label{Z2symm}}

We now show that there exists a finite region in moduli space around the
semi-classical expansion point such that the flux configuration $N_{1}=-N_{2}$
admits critical points satisfying $S_{1}=-\overline{S_{2}}$. \ It follows from
the explicit expressions for the $\tau_{ij}$ given in appendix A that near the
semi-classical expansion point and along the locus $S_{1}=-\overline{S_{2}}$:%
\begin{equation}
\left[
\begin{array}
[c]{cc}%
\overline{\tau_{11}} & \overline{\tau_{12}}\\
\overline{\tau_{12}} & \overline{\tau_{22}}%
\end{array}
\right]  =\left[
\begin{array}
[c]{cc}%
-\tau_{22} & -\tau_{12}\\
-\tau_{12} & -\tau_{11}%
\end{array}
\right]  -\frac{M-\overline{M}}{2\pi i}\left[
\begin{array}
[c]{cc}%
1 & 1\\
1 & 1
\end{array}
\right]  \label{TAUMATrelation}%
\end{equation}
where $M\equiv\log\left(  \Lambda_{0}^{2}/\Delta^{2}\right)  $ and:%
\begin{equation}
\mathcal{F}_{111}=\overline{\mathcal{F}_{222}}\text{,\ \ \ \ }\mathcal{F}%
_{112}=\overline{\mathcal{F}_{122}}. \label{specialthree}%
\end{equation}
Note that the above relations do not require $\Lambda_{0}$ to be a real number.

We now apply the above relations in order to simplify the attractor-like
equations (\ref{firstform}) and (\ref{alpha}). \ Equation (\ref{specialthree})
implies that along this locus, the determinants of equation (\ref{dets})
satisfy $d_{1}=\overline{d_{2}}$ and $d_{3}=\overline{d_{3}}$ so that the
discriminant $d_{3}^{2}-4d_{1}d_{2}$ is a real number. \ It thus follows from
the formulae in appendix A that near the semi-classical expansion point:%
\begin{equation}
d_{3}^{2}-4d_{1}d_{2}>0\text{.}%
\end{equation}
This in turn implies:%
\begin{equation}
\overline{\rho_{v}}=\rho_{w}\text{.} \label{VWrelation}%
\end{equation}
Substituting equations (\ref{TAUMATrelation}) and (\ref{VWrelation}) into the
attractor-like equations (\ref{firstform}) and (\ref{alpha}) yields:%
\begin{align}
\frac{2i}{C^{\prime}}\left[
\begin{array}
[c]{c}%
N_{1}\\
N_{2}%
\end{array}
\right]   &  =\left(  1-\rho_{w}\right)  \left[
\begin{array}
[c]{c}%
1\\
-1
\end{array}
\right] \label{Nequations}\\
\frac{2i}{C^{\prime}}\alpha &  =\left(  \tau_{22}+\tau_{12}+\rho_{w}(\tau
_{11}+\tau_{12})+\frac{M-\overline{M}}{2\pi i}(1+\rho_{w})\right)
\end{align}
where $C^{\prime}$ is a common rescaling factor. \ Hence, $N_{1}=-N_{2}$ and:%
\begin{equation}
\frac{\alpha}{N_{1}}=\frac{1+\rho_{w}}{1-\rho_{w}}\left(  \tau_{11}+\tau
_{12}+\frac{M-\overline{M}}{2\pi i}\right)  +\frac{\tau_{22}-\tau_{11}}%
{1-\rho_{w}}. \label{ANRAT}%
\end{equation}

\subsection{Flux Configurations of the Real Locus\label{reallocus}}

When $S_{1}>0>S_{2}$, an argument similar to the one given in the previous
subsection establishes that in a finite neighborhood of the semi-classical
expansion point, both $\rho_{v}$ and $\rho_{w}$ of equation (\ref{nuequation})
and (\ref{omegaequation}) are again real. \ In this case $\overline{\tau_{ij}%
}=-\tau_{ij}-(M-\overline{M})/2\pi i$ and the attractor-like equations may be
written as:%
\begin{align}
\frac{2i}{C}\left[
\begin{array}
[c]{c}%
N_{1}\\
N_{2}%
\end{array}
\right]   &  =\left[
\begin{array}
[c]{c}%
2\rho_{w}\left(  \tau_{11}-\tau_{12}\right)  +\left(  1+\rho_{v}\rho
_{w}\right)  \left(  \tau_{12}-\tau_{22}\right) \\
2\rho_{v}\left(  \tau_{12}-\tau_{22}\right)  +\left(  1+\rho_{v}\rho
_{w}\right)  \left(  \tau_{11}-\tau_{12}\right)
\end{array}
\right] \\
\frac{2i}{C}\alpha &  =\left(  \rho_{v}\rho_{w}-1\right)  \det\tau
+\frac{M-\overline{M}}{2\pi i}\left(  1+\rho_{v}\right)  \left(  \tau
_{12}-\tau_{22}+\rho_{w}\left(  \tau_{11}-\tau_{12}\right)  \right)  .
\end{align}
Because $\tau_{ij}-\tau_{12}$ is pure imaginary and $\rho_{v}$ and $\rho_{w}$
are purely real, we conclude that $N_{1}/N_{2}$ is purely real. \ Further,
when $M=\overline{M}$, the ratio $\alpha/N_{1}$ is pure imaginary. \ We
therefore conclude that for $\theta_{\mathrm{YM}}=0$ and $M=\overline{M}$,
$V_{\mathrm{eff}}$ admits critical points corresponding to geometries with the
branch cuts aligned along the real axis of the $x$-plane.

\section{Confining Vacua and Glueball Phase Alignment\label{confiningvacua}}

In this section we show that two loop corrections to our metastable
brane/anti-brane system generate an energetically preferred confining vacuum.
\ In the closed string dual this preferred vacuum corresponds to
a\ configuration where the branch cuts align along a common axis. \ We next
estimate the tunneling rate for glueball phase re-alignment and find the rate
of decay to this lowest energy configuration is suppressed by large $N$
effects which can only be countered by an exponentially small glueball field.
\ Viewing our construction as imbedded inside a compact Calabi-Yau, string
theory requires that $\theta_{\mathrm{YM}}$ be treated as a dynamical field.
\ We show that the same effect which lifts the degeneracy between the
confining vacua generates a potential for $\theta_{\mathrm{YM}}$ which is
consistent with general expectations from large $N$ gauge theories.

\subsection{Degenerate Confining Vacua}

It is well-known that pure super Yang-Mills theory with gauge group $SU(N)$
has $N$ confining vacua counted by the Witten index ${\mathrm{Tr}}\left(
-1\right)  ^{F}=N$. \ Indeed, the glueball field of the theory attains $N$
distinct values:%
\begin{equation}
S=\zeta\Lambda^{3} \label{confiningscale}%
\end{equation}
where $\zeta$ denotes an $N^{\mathrm{th}}$ root of unity and $\Lambda$ is the
holomorphic scale of confinement for the gauge group $SU(N)$.

We now show that at \textit{leading order} in the expansion of the periods,
the confining vacua of the $n$ cut geometry are also energetically degenerate.
\ To this end, note that equation (\ref{leadingordertaus}) implies that for
$i\neq j$, $\tau_{ij}$ is constant. \ The claimed degeneracy now follows
because the critical points of $V_{\mathrm{eff}}$ are given by extremizing
with respect to the variables $N_{i}\log\left(  S_{i}/\left(  W^{\prime\prime
}(a_{i})\Lambda_{0}^{2}\right)  \right)  $ for all $i$.

But whereas the energy of each confining vacuum in the supersymmetric case is
zero to \textit{all} orders in an expansion of the periods, higher order
corrections in the non-supersymmetric case should lift this degeneracy.
\ Indeed, our expectation is that the Coulomb attraction between branes and
anti-branes will cause the branch cuts of the closed string dual to align in
order to more efficiently annihilate flux lines. \ We now confirm this in the
case of the two cut geometry.

\subsection{Higher Order Corrections in the Two Cut Geometry}

At leading order, the energy density of the brane/anti-brane system is
\cite{ABSV}:%
\begin{equation}
E^{(0)}=\frac{8\pi}{g_{\mathrm{YM}}^{2}}\left(  \left\vert N_{1}\right\vert
+\left\vert N_{2}\right\vert \right)  -\frac{2}{\pi}\left\vert N_{1}%
\right\vert \left\vert N_{2}\right\vert \log\left\vert \frac{\Lambda_{0}%
}{\Delta}\right\vert ^{2}\text{.} \label{energyleading}%
\end{equation}
Because $E^{(0)}$ does not depend on the phases of the glueball fields, the
confining vacua are energetically degenerate.

At higher order the effective potential takes the form:%
\begin{equation}
V_{\mathrm{eff}}=V_{\mathrm{eff}}^{(0)}+V_{\mathrm{pert}}\text{.}%
\end{equation}
Incorporating the two loop correction to $\tau_{ij}$ given in appendix A lifts
the degeneracy in energy densities:%
\begin{equation}
E=E^{(0)}-\frac{10\left\vert N_{1}\right\vert \left\vert N_{2}\right\vert
}{\pi}\left(  t_{1}+\overline{t_{1}}+t_{2}+\overline{t_{2}}\right)  ,
\label{energyaxions}%
\end{equation}
where $t_{1}\equiv S_{1}/(g\Delta^{3})$ and $t_{2}\equiv-S_{2}/(g\Delta^{3})$
are given by equations (\ref{leadstwo}) and (\ref{leadsthree}), respectively.
\ The confining vacuum with the lowest energy density is given by the
configuration with $t_{1}$ and $t_{2}$ as close to being real positive numbers
as possible. \ Without loss of generality, the geometrical significance of
this result can be seen when $\Delta>0$. \ It now follows from equations
(\ref{sone}) and (\ref{stwo}) and the remarks below these equations that in
this case the branch cuts are nearly aligned along the line joining $x=a_{1}$
and $x=a_{2}$.

\subsection{Tunneling Rates}

In the previous subsection we showed that the two loop contribution to
$V_{\mathrm{eff}}$ lifts the degeneracy in energy density between the many
confining vacua of the theory. \ In this subsection we compute the tunneling
rate for glueball phase re-alignment. \ In particular, we find that in the
strict $N_{i}=\infty$ limit with fixed confinement scale, these additional
confining vacua become exactly stable. \ For finite $N_{i}$, this leads to the
presence of a large number of additional metastable vacua of the type
generically present in large $N$ gauge theories
\cite{WittenthetaSolution,ShifmanTunneling}. \ On the other hand, treating the
glueball field as an exponentially suppressed quantity, we find that the
corresponding tunneling rate increases.

In general, such tunneling events correspond to the nucleation of a bubble of
vacuum with lower energy density inside the higher energy density vacuum.
\ Assuming that in the Euclidean continuation of Minkowski space that this
bubble is an $O(4)$ symmetric configuration, the thin wall approximation of
the tunneling rate is \cite{ColemanTunnel}:%
\begin{equation}
\Gamma\sim\exp\left(  -S_{bounce}\right)  =\exp\left(  -\frac{27\pi^{2}}%
{2}\frac{T^{4}}{\left(  \Delta V\right)  ^{3}}\right)
\label{tunnrateconfining}%
\end{equation}
where $T$ is the tension of the domain wall and $\Delta V$ is the change in
energy density between the two vacua.

The domain wall solutions separating the confining vacua of the theory are
given by wrapping D5-branes over the $A$-cycle threaded by positive flux and
anti-D5-branes over the $A$-cycle threaded by negative flux. \ We approximate
the tension of such a domain wall using the supersymmetric analogue with
$N_{1},N_{2}>0$:%
\begin{equation}
T=\frac{1}{g_{s}}\left\vert \mathcal{W}_{\mathrm{eff}}(\zeta_{1}S_{1}%
,\zeta_{2}S_{2})-\mathcal{W}_{\mathrm{eff}}(\zeta_{1}^{\prime}S_{1},\zeta
_{2}^{\prime}S_{2})\right\vert
\end{equation}
where $\zeta_{i}$ and $\zeta_{i}^{\prime}$ are $N_{i}^{\mathrm{th}}$ roots of
unity and the $S_{i}$'s are evaluated at a supersymmetric critical
point\footnote{The additional factor of $1/g_{s}$ in the tension formula
follows from the string frame normalization of the superpotential described in
the footnote above equation (\ref{Veffdef}).}. \ For simplicity, we now
restrict our analysis to tunneling events which only rephase $S_{1}$. \ The
leading order $S_{1}$ dependence of $\Pi_{1}$ and $\Pi_{2}$ is:%
\begin{align}
2\pi i\Pi_{1}  &  =S_{1}\left(  \log\frac{S_{1}}{g\Delta\Lambda_{0}^{2}%
}-1\right) \\
2\pi i\Pi_{2}  &  =S_{1}\log\frac{\Delta^{2}}{\Lambda_{0}^{2}}%
\end{align}
The tension of the domain wall solution which interpolates between different
discrete phase choices for $S_{1}$ is therefore:%
\begin{equation}
T=\frac{1}{2\pi g_{s}}\left\vert N_{1}S_{1}\left(  \zeta_{1}-\zeta_{1}%
^{\prime}\right)  \right\vert =\frac{1}{\pi g_{s}}\left\vert N_{1}S_{1}%
\sin\frac{\pi\left(  l-l^{\prime}\right)  }{N_{1}}\right\vert .
\end{equation}
where $\zeta_{1}=\exp\left(  2\pi il/\left\vert N_{1}\right\vert \right)  $
and $\zeta_{1}^{\prime}=\exp\left(  2\pi il^{\prime}/\left\vert N_{1}%
\right\vert \right)  $. \ The change in energy density between the two vacua
has norm:%
\begin{equation}
\left\vert \Delta V\right\vert =\frac{20\left\vert N_{1}\right\vert \left\vert
N_{2}\right\vert \left\vert t_{1}\right\vert }{\pi}\left\vert \cos\left(
\frac{\widehat{\theta}+2\pi l}{\left\vert N_{1}\right\vert }\right)
-\cos\left(  \frac{\widehat{\theta}+2\pi l^{\prime}}{\left\vert N_{1}%
\right\vert }\right)  \right\vert
\end{equation}
where $\widehat{\theta}/\left\vert N_{1}\right\vert $ denotes the argument of
$t_{1}$.

We now estimate the value of $T^{4}/\left(  \Delta V\right)  ^{3}$ for
different glueball phase alignment tunneling events. \ In particular, we show
that glueball phase alignment typically proceeds via a single large drop in
energy density rather than a sequence of tunneling events with smaller drops
in energy. \ As it is irrelevant for the considerations to follow, we suppress
all dependence on $\widehat{\theta}$ in the expressions to follow.

First consider the tunneling event from $l=N_{1}/2$ to $l^{\prime}=0$
corresponding to a single instanton process with the largest possible drop in
energy density. \ The bounce action is proportional to:%
\begin{equation}
S_{N_{1}/2\rightarrow0}\varpropto T^{4}/\left(  \Delta V\right)  ^{3}=\frac
{1}{40^{3}\pi}\frac{\left\vert N_{1}\right\vert }{\left\vert N_{2}\right\vert
^{3}}\frac{\left\vert t_{1}\right\vert \left\vert g\Delta^{3}\right\vert ^{4}%
}{g_{s}^{4}}\text{.} \label{fulldrop}%
\end{equation}
Because the scaling of the $N_{i}$ and $g_{s}^{-1}$ in the large $N$ limit are
all comparable, note that the corresponding bounce action scales as $N^{2}$ so
that for finite $t_{1}$ this tunneling event is highly suppressed. \ Note,
however, that it is also natural to consider the limit in which $t_{1}$ is
exponentially suppressed. \ In this case the tunneling rate increases.

Next consider the tunneling process from the vacuum $l=N_{1}/2$ to $l^{\prime
}=N_{1}/2-\delta$ for $\delta$ small compared to $N_{1}$. \ In this case, the
bounce action is proportional to:%
\begin{equation}
S_{N_{1}/2\rightarrow N_{1}/2-\delta}\varpropto T^{4}/\left(  \Delta V\right)
^{3}=\frac{1}{40^{3}\pi^{3}}\left\vert \frac{N_{1}}{N_{2}}\right\vert
^{3}\frac{\left\vert t_{1}\right\vert \left\vert g\Delta^{3}\right\vert ^{4}%
}{\delta^{2}g_{s}^{4}} \label{smalldrop}%
\end{equation}
which scales as $N^{4}$ in the large $N$ limit. \ The tunneling rate for a
small re-alignment in the glueball phase is therefore much smaller than that
due to a large re-alignment in phase.

This conclusion is further supported by considering the tunneling process from
a vacuum $l=\delta$ to $l^{\prime}=0$ where $\delta$ is again small compared
to $\left\vert N_{1}\right\vert $. \ In this case we find that the bounce
action is given by the same expression as equation (\ref{smalldrop}). \ The
decay rate due to rephasing $t_{1}$ is therefore:%
\begin{equation}
\Gamma\sim\exp\left(  -\frac{27}{2}\frac{1}{40^{3}\pi}\left\vert \frac{N_{1}%
}{N_{2}}\right\vert ^{3}\frac{\left\vert t_{1}\right\vert \left\vert
g\Delta^{3}\right\vert ^{4}}{\delta^{2}g_{s}^{4}}\right)  \text{.}%
\label{angledecay}%
\end{equation}
It therefore follows that as $\delta$ increases, the corresponding tunneling
rate also increases. \ Tuning the parameters of the theory so that $t_{1}$
remains fixed, note that when the ratio $\left\vert N_{1}/N_{2}\right\vert
\gg$ $1$, the tunneling rate is suppressed. \ Conversely, when the ratio
$\left\vert N_{2}/N_{1}\right\vert \gg$ $1$, the tunneling rate is higher and
the smaller cut aligns along the real axis on a shorter time scale.
\ Physically this corresponds to the fact that the orientation of a cut
fluctuates less as the amount of flux passing through the corresponding
$A$-cycle increases.

\subsection{Axion Potential\label{axions}}

As we have seen above, the two loop contribution to $V_{\mathrm{eff}}$ aligns
the phases of the glueball fields. \ It follows from string theory that upon
imbedding our model in a compact Calabi-Yau threefold, $\theta_{\mathrm{YM}}$
must be treated as a dynamical field\footnote{In a compact Calabi-Yau, the
relative separation $\Delta$ between the branes becomes a normalizable mode.
\ In this section we assume that there exists a mechanism which stabilizes
this value.} with potential given by equation (\ref{energyaxions}) evaluated
at the preferred confining vacuum. \ The effective value of the $\theta$-angle
on the branes and anti-branes follows from the relation:%
\begin{equation}
S_{i}=\zeta_{i}\Lambda_{i}^{3}=\zeta_{i}\left\vert \Lambda_{i}\right\vert
^{3}\exp\left(  i\theta_{i}/\left\vert N_{i}\right\vert \right)
\end{equation}
where $\zeta_{i}$ denotes an $\left\vert N_{i}\right\vert ^{\text{th}}$ root
of unity.

We now illustrate the form of the axion potential in the case $N_{1}%
=-N_{2}\equiv N$. \ Equations (\ref{leadstwo}) and (\ref{leadsthree}) imply
that for such flux configurations, the phase of $\Lambda_{0}/\Delta$ does not
contribute to the phases of the $S_{i}$'s. \ For simplicity, we further
restrict to the case where $g\Delta^{3}$ is purely real. \ In this case, the
effective value of the $\theta$-angles for the branes and anti-branes satisfy
$\theta_{1}=-\theta_{2}=$ $\theta_{\mathrm{YM}}$. \ The energy now takes the
form:%
\begin{equation}
E=E^{(0)}-\frac{20N^{2}\left\vert t\right\vert }{\pi}\left(  \cos\left(
\frac{\theta_{\mathrm{YM}}+2\pi l}{N}\right)  +\cos\left(  \frac
{\theta_{\mathrm{YM}}+2\pi l^{\prime}}{N}\right)  \right)
\end{equation}
where $t\equiv t_{1}=t_{2}$ and $l$ and $l^{\prime}$ are integers.
\ Evaluating $l$ and~$l^{\prime}$ at the preferred confining vacuum
configuration yields a potential for the axion:%
\begin{equation}
V_{\mathrm{ax}}\left(  \theta_{\mathrm{YM}}\right)  =E^{(0)}-\frac
{40N^{2}\left\vert t\right\vert }{\pi}\underset{l\in%
\mathbb{Z}
}{\min}\left(  \cos\left(  \frac{\theta_{\mathrm{YM}}+2\pi l}{N}\right)
\right)  =E^{(0)}-\frac{40N^{2}\left\vert t\right\vert }{\pi}\cos\left(
\frac{\theta_{\mathrm{YM}}}{N}\right)  \label{axionpotential}%
\end{equation}
where in the last equality we have assumed that $\theta_{\mathrm{YM}}%
\in\left[  -\pi,\pi\right)  $. \ This potential has a minimum at
$\theta_{\mathrm{YM}}=0$. \ In the more general case where $\left\vert
N_{1}\right\vert \neq\left\vert N_{2}\right\vert $ and both $\Delta$ and
$g\Delta$ acquire complex phases, the minimum of $V_{\mathrm{ax}}$ will be
shifted away from this value. \ Naively, the requirement that the physics
remain invariant under the substitution $\theta_{\mathrm{YM}}\rightarrow
\theta_{\mathrm{YM}}+2\pi$ appears incompatible with the $\theta_{\mathrm{YM}%
}/N$ dependence of the final expression of equation (\ref{axionpotential}).
\ That the physics does remain invariant follows from the first equality of
equation (\ref{axionpotential}).

In fact, this is a general issue in large $N$ gauge theories. \ Recall that in
large $N$ $U(N)$ pure Yang-Mills theory, the Lagrangian density is:%
\begin{equation}
\mathcal{L}=-\frac{N}{4\lambda}Tr\left(  F_{\mu\nu}F^{\mu\nu}\right)
+\frac{\theta}{64\pi^{2}}\varepsilon^{\alpha\beta\mu\nu}Tr\left(
F_{\alpha\beta}F_{\mu\nu}\right)
\end{equation}
where $\lambda$ denotes the 't Hooft coupling of the theory. \ The dependence
of the vacuum energy density on the $\theta$-angle is:%
\begin{equation}
E_{vac}\left(  \theta\right)  =N^{2}f\left(  \frac{\theta}{N}\right)
\label{largeNenergy}%
\end{equation}
where $f$ is a function which is well-defined in the large $N$ limit. \ The
factor of $N^{2}$ arises from the number of degrees of freedom and the
$\theta/N$ dependence follows from the requirement that the energy density
must possess a well-defined large $N$ limit
\cite{WittenLargeNUONEproblem,WittenLargeNChirDyn,WittenthetaSolution}. $\ $In
order to restore invariance under $\theta\rightarrow\theta+2\pi$, it is
natural to conjecture that the full $\theta$ dependence of $E_{vac}$ is
\cite{WittenLargeNChirDyn}:%
\begin{equation}
E_{vac}\left(  \theta\right)  =N^{2}\underset{l\in%
\mathbb{Z}
}{\min}f\left(  \frac{\theta+2\pi l}{N}\right)  \label{evacbranches}%
\end{equation}
where the integers $l$ label distinct metastable branches of vacua. \ It has
been shown that this type of vacuum structure arises both in softly broken
supersymmetric QCD with small gaugino masses
\cite{ShifmanthetaSolution,EvansthetaSolution} as well as from D-brane
constructions of large $N$ gauge theories \cite{WittenthetaSolution}. \ In the
limit $N=\infty$, it is expected that if $t_{1}$ remains fixed and in
particular is not exponentially suppressed, that these additional vacua become
exactly stable \cite{WittenthetaSolution,ShifmanTunneling}. \ This is indeed
consistent with the decay rates given by equations (\ref{fulldrop}),
(\ref{smalldrop}) and (\ref{angledecay}). \ Multiple branches of vacua may
also be present in QCD
\cite{HalpZhitTopSusc,HalpZhitCanTheta,HalpZhitIntegratingIn,HalpZhitAnomalous,ZhitDefects}%
.

Returning to the first equality of equation (\ref{axionpotential}), note that
our expression for the vacuum energy density is of \textit{precisely} the form
conjectured in equation (\ref{evacbranches}). \ This provides an exactly
calculable example where the expectations discussed above are explicitly borne
out. \ Indeed, although there are higher order corrections to $E_{vac}$ and
therefore to $V_{\mathrm{ax}}$, to leading order in $1/N$, \textit{all} of
these corrections are captured by the closed string dual description.

\section{Breakdown of Metastability\label{Breakdown}}

When the amount of flux through each $S^{3}$ is sufficiently low, the
corresponding glueball field is stabilized at a small value. \ But as shown in
section \ref{confiningvacua}, two loop contributions to $V_{\mathrm{eff}}$
\ generate a preferred confining vacuum which aligns the phases of the
glueball fields. \ In this section we show that for sufficiently large values
of the 't Hooft coupling, this same two loop effect lifts the metastable vacua
present at weak coupling. \ Note in particular that because the corrections to
the K\"{a}hler potential are of order $1/N$, the holographic dual description
of the brane dynamics in terms of the flux induced potential $V_{\mathrm{eff}%
}$ becomes \textit{more} accurate as the amount of flux through each $S^{3}$ increases.

Before proceeding with a more precise analysis, we first give a heuristic
derivation of the value of the 't Hooft coupling for which we expect higher
order corrections to $V_{\mathrm{eff}}$ to lift the metastable vacua present
at weak coupling. \ Recall from equation (\ref{energyleading}) that the
leading order energy density of the brane/anti-brane system is:%
\begin{equation}
E^{(0)}=\frac{8\pi}{g_{\mathrm{YM}}^{2}}\left(  \left\vert N_{1}\right\vert
+\left\vert N_{2}\right\vert \right)  -\frac{2}{\pi}\left\vert N_{1}%
\right\vert \left\vert N_{2}\right\vert \log\left\vert \frac{\Lambda_{0}%
}{\Delta}\right\vert ^{2}\text{.}%
\end{equation}
The first term corresponds to the bare tension of the branes and the second
term corresponds to the Coulomb attraction between the branes.

Returning to the discussion near equation (\ref{shiftedpotential}), it follows
that when $\left\vert N_{1}\right\vert \gtrsim\left\vert N_{2}\right\vert $
and $N_{1}>0>N_{2}$, we have:
\begin{equation}
E^{(0)}\geq\frac{8\pi}{g_{\mathrm{YM}}^{2}}\left(  N_{1}+N_{2}\right)
\label{ebounded}%
\end{equation}
with similar inequalities for different choices of relative magnitudes and
signs for the $N_{i}$. \ We expect to lose metastability precisely when the
Coulomb attraction contribution to the energy density becomes comparable to
the bare tension of the branes. \ This is near the regime where $E^{(0)}$ is
close to saturating inequality (\ref{ebounded}). \ This yields the following
estimate for the breakdown of metastability:
\begin{equation}
\frac{1}{g_{\mathrm{YM}}^{2}\left\vert N_{1}\right\vert }\sim\log\left\vert
\frac{\Lambda_{0}}{\Delta}\right\vert ^{2} \label{firstestimate}%
\end{equation}
where in the above expression we have dropped all factors of order unity.

Perhaps more surprisingly, this breakdown in metastability is calculable near
the semi-classical expansion point. \ Indeed, for illustrative purposes we
show in the subsection to follow that for $N_{1}=-N_{2}$, the two loop
contribution to the glueball potential causes $V_{\mathrm{eff}}$ to develop a
local maximum. \ Beyond this local maximum the potential subsequently rolls
downward beyond the regime where perturbations about the semi-classical
expansion point provide an accurate description.

The rest of this section is organized as follows. \ For simplicity and with
the analysis of section \ref{confiningvacua} in mind, we restrict to flux
configurations which produce metastable vacua with the branch cuts aligned
along the real axis of the complex $x$-plane. \ In subsection \ref{twolooper}
we illustrate in the case $N_{1}=-N_{2}$ that the two loop contribution to
$V_{\mathrm{eff}}$ generates a local maximum for the glueball potential. \ In
subsection \ref{Z2breakdown} we study the breakdown in metastability for flux
configurations with $N_{1}=-N_{2}$ and $\theta_{\mathrm{YM}}=0$, and in
subsection \ref{ONELARGE} we perform a similar analysis for the case
$\left\vert N_{1}\right\vert \gg\left\vert N_{2}\right\vert $ and
$\theta_{\mathrm{YM}}=0$. \ In both cases we find that the value of the flux
at which metastability breaks down is in rough agreement with the estimate
given by equation (\ref{firstestimate}).

\subsection{Two Loop Corrections to $V_{\mathrm{eff}}$\label{twolooper}}

In this subsection we show that the higher order corrections to the periods
alter the shape of the flux induced effective potential. \ Along the locus
$S_{1}=-S_{2}>0$, the leading order behavior of the effective potential is:%
\begin{equation}
V_{\mathrm{eff}}^{(0)}=\frac{\left\vert \alpha\right\vert ^{2}}{\pi\left(
2M-\log t\right)  }\left(  4\pi^{2}-\left\vert \frac{N}{\alpha}\right\vert
^{2}\left(  2M-\log t\right)  \log t\right)
\end{equation}
where we have introduced the parameter $t\equiv S_{1}/g\Delta^{3}%
=-S_{2}/g\Delta^{3}$. \ This potential possesses a single extremum which is a
minimum. \ We now show that this behavior is only correct for very small $t$.
\ The higher order behavior of $V_{\mathrm{eff}}$ is:%
\begin{equation}
V_{\mathrm{eff}}=\frac{\left\vert \alpha\right\vert ^{2}}{\pi\left(
2M+6t-\log t\right)  }\left(
\begin{array}
[c]{c}%
4\pi^{2}-\left\vert \frac{N}{\alpha}\right\vert ^{2}\left(  2M-\log t\right)
\log t\\
-\left\vert \frac{N}{\alpha}\right\vert ^{2}\left(  68Mt+204t^{2}+28t\log
t\right)
\end{array}
\right)  .
\end{equation}
To determine the appearance of a local maximum in $V_{\mathrm{eff}}$, we
compare the relative $t$ dependence of all terms in the above expression.
\ The appearance of a local maximum is due to the term proportional to $68Mt$.
\ We may therefore approximate $V_{\mathrm{eff}}$ as:%
\begin{align}
V_{\mathrm{eff}}  &  \simeq V_{\mathrm{eff}}^{(0)}-\frac{68M\left\vert
N\right\vert ^{2}}{\pi}\frac{t}{2M-\log t}\\
&  \equiv V_{\mathrm{eff}}^{(0)}+V^{(2)}\text{.}%
\end{align}
We now show that for a suitable range of values, this new term causes
$V_{\mathrm{eff}}$ to develop a local maximum and a local minimum. \ 

To this end, first consider the derivative of $V_{\mathrm{eff}}^{(0)}$ with
respect to $t$:%
\begin{equation}
\frac{dV_{\mathrm{eff}}^{(0)}}{dt}\simeq\frac{\left\vert \alpha\right\vert
^{2}}{\pi t(2M-\log t)^{2}}\left(  -4M^{2}\left\vert \frac{N}{\alpha
}\right\vert ^{2}+4\pi^{2}+4M\left\vert \frac{N}{\alpha}\right\vert ^{2}\log
t-\left\vert \frac{N}{\alpha}\right\vert ^{2}(\log t)^{2}\right)  .
\end{equation}
Note that the term in parentheses is a quadratic polynomial in $\log t$ and
therefore achieves a maximal value. \ By including the contribution from
$dV^{(2)}/dt$, we find that for suitable values of $\left\vert N/\alpha
\right\vert $, the equation $dV_{\mathrm{eff}}/dt=0$ possesses at least two
solutions. \ These solutions correspond to a metastable minimum and a nearby
maximum for the effective potential.%
\begin{figure}
[ptb]
\begin{center}
\includegraphics[
height=2.479in,
width=4.005in
]%
{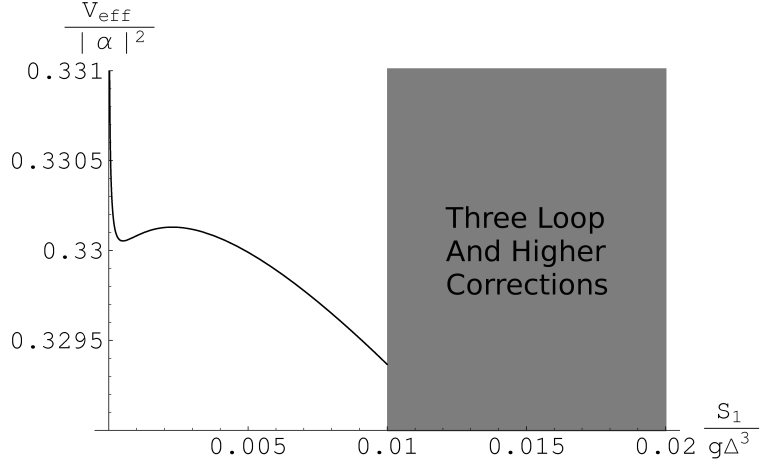}%
\caption{Plot of $V_{\mathrm{eff}}/\left\vert \alpha\right\vert ^{2}$ in the
two loop approximation along the locus $S_{1}/g\Delta^{3}=-S_{2}/g\Delta
^{3}>0$ for a flux configuration with $N_{1}=-N_{2}$. \ In this plot
$\Lambda_{0}/\Delta\sim10^{4}$ and $\left\vert N_{1}/\alpha\right\vert
\sim0.1$.}%
\label{TWOLOOPPLOT}%
\end{center}
\end{figure}

\subsection{Breakdown of Metastability: $N_{1}=-N_{2}$\label{Z2breakdown}}

When $N_{1}=-N_{2}\equiv N$ and $\theta_{\mathrm{YM}}=0$, the metastable
minima of $V_{\mathrm{eff}}$ correspond to two equal size branch cuts aligned
along the real axis of the complex $x$-plane. \ To study the appearance of an
instability in $V_{\mathrm{eff}}$, we return to our general approach of
solving for the fluxes as a function of moduli. \ The only non-trivial moduli
dependence arises from the ratio:%
\begin{equation}
\frac{\alpha}{N}=-\frac{4\pi i}{g_{\mathrm{YM}}^{2}N}.
\end{equation}
Along the locus $S_{1}=-S_{2}>0$, the moduli dependent function $\alpha/N$ may
assume the same value multiple times. \ This implies that for a given set of
fluxes, there are multiple critical points of $V_{\mathrm{eff}}$. \ 

We now briefly switch perspectives and view these critical points as functions
of $\alpha/N$. \ When these critical points approach the same point in moduli
space, the effective potential develops a flat direction. \ Viewed as a
function of moduli, when the 't Hooft parameter:%
\begin{equation}
\lambda\equiv g_{\mathrm{YM}}^{2}\left\vert N\right\vert
\end{equation}
reaches a maximal value, the system develops an instability. \ 

We now determine the value of $t$ for which $\lambda$ attains a maximum. \ It
follows from equation (\ref{ANRAT}) and the expressions of appendix A that:%
\begin{equation}
\frac{8\pi^{2}}{\lambda}=-\log t+M+\overline{M}+t(6+20M+20\overline{M}-10\log
t^{2})+O(t^{2}\log t)\text{.} \label{ANRATEXPANDED}%
\end{equation}
Note that to leading order in $t$, $\lambda^{-1}\sim$ $-\log t+M+\overline{M}$
and therefore does not possess a minimum. \ The quantity $\lambda^{-1}$ is
minimized when:%
\begin{equation}
t_{\ast}=-\frac{1}{20W_{-1}\left(  -e^{-L}\right)  } \label{etasolved}%
\end{equation}
where $L\equiv\log\left\vert \Lambda_{0}^{4}/\Delta^{4}\right\vert
+\log\left(  20e^{-7/10}\right)  $ and $W_{-1}$ is the $-1^{\mathrm{th}}$
branch of the Lambert W-function.

It is also of interest to study the $\Delta$ dependence of $t_{\ast}$. \ For
sufficiently small $t_{\ast}$ and $\Delta$, we have:%
\begin{equation}
\log\left\vert \frac{\Lambda_{0}}{\Delta}\right\vert ^{4}\simeq\frac
{1}{20t_{\ast}}\text{.} \label{rstar}%
\end{equation}
Fixing the value of $t$, we conclude that the system always develops an
instability once $\Delta$ is sufficiently small. \ In particular, we see that
by tuning $\Delta$ to a sufficiently small value, $t_{\ast}$ can be made
arbitrarily small, justifying the two loop approximation.

To conclude this subsection we estimate the value of $g_{\mathrm{YM}}%
^{2}\left\vert N\right\vert $ for which $V_{\mathrm{eff}}$ develops an
instability. \ It follows from equations (\ref{ANRATEXPANDED}) and
(\ref{rstar}) that for sufficiently small $\Delta$ and $t_{\ast}$:%
\begin{equation}
\exp\left(  \frac{8\pi^{2}}{\lambda_{\ast}}\right)  =20\left\vert
\frac{\Lambda_{0}}{\Delta}\right\vert ^{4}\log\left\vert \frac{\Lambda_{0}%
}{\Delta}\right\vert ^{4} \label{refined}%
\end{equation}
where $\lambda_{\ast}$ is the value of the bare 't Hooft coupling at the phase
transition. \ If we now drop the factors $20$ and $\log\left\vert
\frac{\Lambda_{0}}{\Delta}\right\vert ^{4}$ in the above expression, we
obtain:%
\begin{equation}
\frac{1}{\lambda_{\ast}}\sim\frac{1}{4\pi^{2}}\log\left\vert \frac{\Lambda
_{0}}{\Delta}\right\vert ^{2} \label{breakdown}%
\end{equation}
up to factors of order unity. \ This is in agreement with the analysis leading
to equation (\ref{firstestimate}). \ Even so, equation (\ref{breakdown})
should be viewed as a crude upper bound.

\subsubsection{Masses and the Mode of Instability: $N_{1}=-N_{2}$}

\begin{figure}
[ptb]
\begin{center}
\includegraphics[
height=2.1835in,
width=5.2038in
]%
{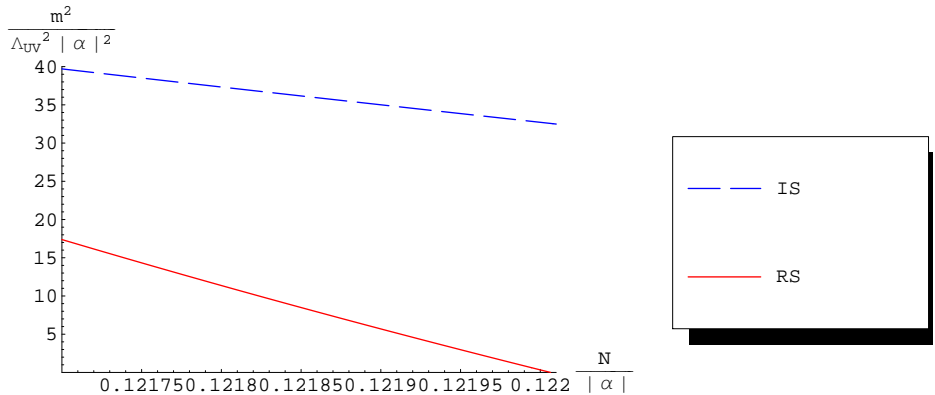}%
\caption{Plot of the two smallest masses $m_{RS}^{2}$ and $m_{IS}^{2}$ as a
function of $\left\vert N/\alpha\right\vert $ for $\Lambda_{0}/\Delta
\sim10^{4}$. \ At a value of $\left\vert N/\alpha\right\vert \sim0.122$ the
system develops an instability. \ This value is in rough agreement with the
estimate of equation (\ref{breakdown}).}%
\label{LINMASS}%
\end{center}
\end{figure}

Although the above analysis establishes that $V_{\mathrm{eff}}$ develops an
instability for sufficiently large values of the flux, it does not directly
indicate the mode of instability. \ It also does not address whether
additional modes of instability appear before reaching such a flux
configuration.\ \ We now show that all the other modes are stable up to this
point and that the unstable mode of the system corresponds to the cuts
remaining equal in size and expanding towards each other. \ To establish this,
we now compute the bosonic mass spectrum for fluxes close to the value where
we expect to lose metastability. \ After canonically normalizing the kinetic
terms of $S_{1}$ and $S_{2}$ and expanding $V_{\mathrm{eff}}$ to quadratic
order, the masses squared are:%
\begin{align}
m_{{RA}}^{2}  &  =\frac{a^{2}}{1-v}+2a\left\vert N\right\vert \left(
-\frac{10}{1+\sqrt{v}}+\frac{7}{(1-v)\pi\operatorname{Im}\tau_{11}}\right)
\label{RA}\\
m_{{RS}}^{2}  &  =\frac{a^{2}}{1-v}+2a\left\vert N\right\vert \left(
-\frac{10}{1-\sqrt{v}}+\frac{7}{(1-v)\pi\operatorname{Im}\tau_{11}}\right) \\
m_{{IS}}^{2}  &  =\frac{a^{2}}{\left(  1+\sqrt{v}\right)  ^{2}}+2a\left\vert
N\right\vert \left(  \frac{10}{1+\sqrt{v}}+\frac{-3}{\left(  1+\sqrt
{v}\right)  ^{2}\pi\operatorname{Im}\tau_{11}}\right) \\
m_{{IA}}^{2}  &  =\frac{a^{2}}{\left(  1-\sqrt{v}\right)  ^{2}}+2a\left\vert
N\right\vert \left(  \frac{10}{1-\sqrt{v}}+\frac{17}{\left(  1-\sqrt
{v}\right)  ^{2}\pi\operatorname{Im}\tau_{11}}\right)  \label{IA}%
\end{align}
where in the above, $RA$ denotes the real anti-symmetric mode corresponding to
both $S_{i}$'s real with one cut growing while the other shrinks, $RS$ denotes
the real symmetric mode corresponding to both $S_{i}$'s real with both cuts
growing in size together, and $IS$ and $IA$ are similarly defined for the
imaginary components of the $S_{i}$'s. \ Further, we have introduced the
parameters:%
\begin{equation}
a=\frac{|N|}{2\pi t\operatorname{Im}{\tau}_{11}}\text{, }v=\frac
{\operatorname{Im}{{\tau}_{12}}^{2}}{\operatorname{Im}{\tau}_{11}%
\operatorname{Im}{\tau}_{22}}%
\end{equation}
where all moduli dependent functions are explicitly evaluated at the critical
point determined by the 't Hooft coupling $\lambda$. \ In equations
(\ref{RA}-\ref{IA}), the term proportional to $a^{2}$ corresponds to the
leading order contribution to the masses squared computed in \cite{ABSV}, and
the term proportional to $2a\left\vert N\right\vert $ corresponds to the two
loop correction to this value. \ As expected based on general symmetry
arguments, we find that as a function of $\left\vert N/\alpha\right\vert $,
$m_{RS}^{2}$ approaches zero as the flux approaches the value given by
equation (\ref{breakdown}). \ In {figure (\ref{LINMASS})} we show the behavior
of $m_{RS}^{2}$ and the next smallest mass squared $m_{IS}^{2}$ as a function
of $\left\vert N/\alpha\right\vert $ near the regime where metastability is
lost.%

\begin{figure}
[ptb]
\begin{center}
\includegraphics[
height=2.4956in,
width=5.2204in
]%
{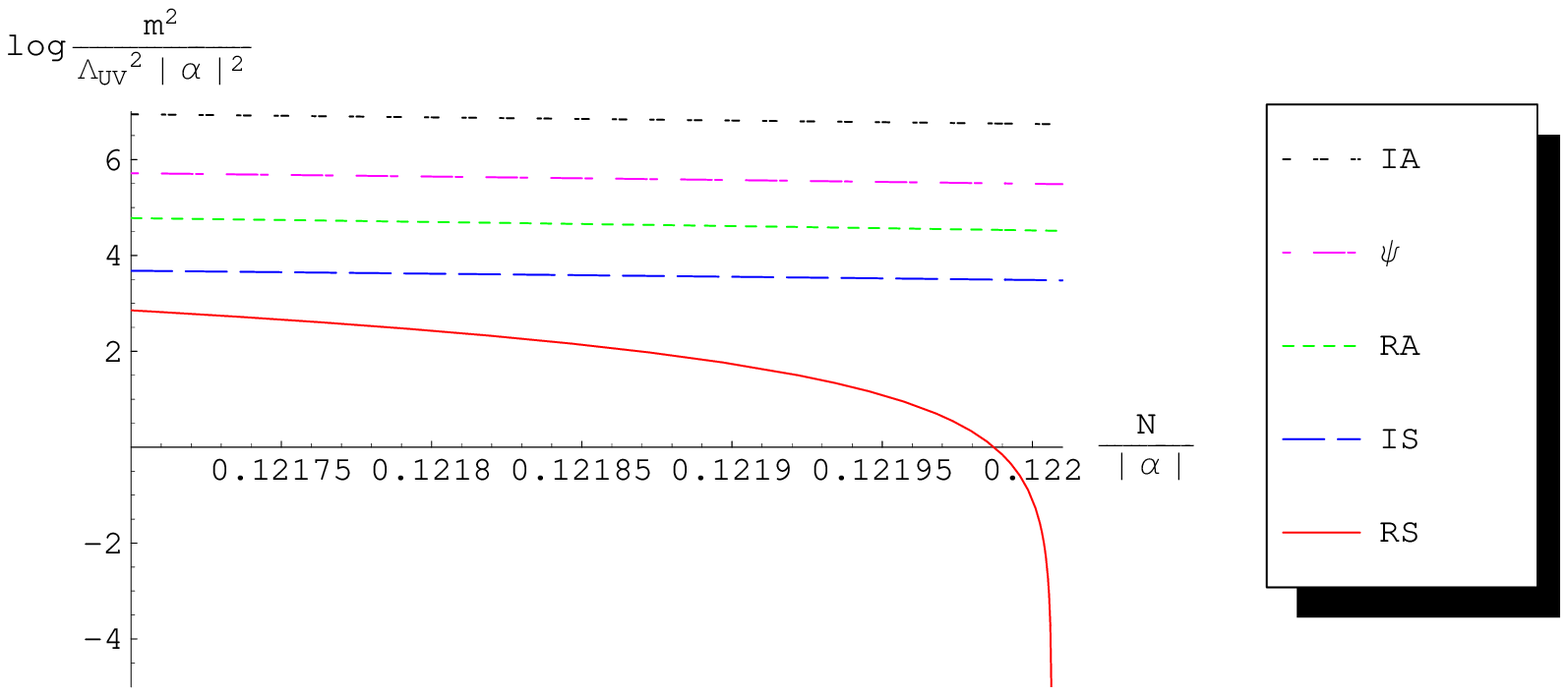}%
\caption{Plot of $\log(m^{2}/(\left\vert \alpha\right\vert ^{2}\Lambda
_{UV}^{2}))$ as a function of $\left\vert N/\alpha\right\vert $ for
$\Lambda_{0}/\Delta\sim10^{4}$. $\ $The bosonic and fermionic masses squared
are given by equations (\ref{RA}-\ref{IA}) and (\ref{fermionmass}),
respectively.}%
\label{LOGMASS}%
\end{center}
\end{figure}

It is also of interest to consider the difference in masses between the
bosonic and fermionic fluctuations dictated by the underlying $\mathcal{N}=2$
structure of the theory. \ We find that the masses of the fermions naturally
group into two sets of values. \ At leading order in $1/N$, the $\mathcal{N}%
=2$ supersymmetry of the theory is spontaneously broken. \ This indicates the
presence of two massless goldstinos. \ Labeling the fermionic counterparts of
the gauge bosons and the $S_{i}$'s respectively by $\psi_{A}^{(i)}$ and
$\psi_{S}^{(i)}$, we find that when $N_{1}=-N_{2}$, the non-zero masses of the
canonically normalized fermionic fields are all equal and given by the value:%
\begin{equation}
|m_{\psi}|=\frac{a}{\left(  1-v\right)  }+\left\vert N\right\vert
\frac{7+10\sqrt{v}}{1-v} \label{fermionmass}%
\end{equation}
where as before, the first term corresponds to the leading order mass and the
second term is the two loop correction to this value. \ In {figure
(\ref{LOGMASS})} we compare the masses squared of the bosonic and fermionic
fluctuations.%

Although we do not include the details here, we find more generally that for
vacua which satisfy $S_{1}=-\overline{S_{2}}$, the system develops an
instability at a similar value of $\left\vert N/\alpha\right\vert $. \ In this
case, the mode of instability causes the cuts to expand in size and rotate
towards the real axis of the complex $x$-plane. \ This is in agreement with
the physical expectation that the flux lines annihilate most efficiently when
the branch cuts are aligned along the real axis of the complex $x$-plane.

\subsection{Breakdown of Metastability: $\left\vert N_{1}\right\vert
\gg\left\vert N_{2}\right\vert $\label{ONELARGE}}

We now study the behavior of $V_{\mathrm{eff}}$ for flux configurations with
$\left\vert N_{1}\right\vert \gg\left\vert N_{2}\right\vert $ and
$\theta_{\mathrm{YM}}=0$. \ More precisely, we also take $N_{1}$ small enough
so that the two loop approximation of $V_{\mathrm{eff}}$ is valid. \ In this
case, the modulus $t_{1}\equiv S_{1}/g\Delta^{3}$ fluctuates much less than
$t_{2}\equiv-S_{2}/g\Delta^{3}$. \ Further, because the behavior of
$V_{\mathrm{eff}}$ is relatively insensitive to the value of $\alpha/N_{2}$,
it is sufficient to fix $\alpha/N_{2}$ and determine the value of
$\alpha/N_{1}$ for which $V_{\mathrm{eff}}$ develops a flat direction.

To this end, we employ a strategy similar to that of subsection
\ref{Z2breakdown} and\ use the attractor-like equations to treat $\alpha
/N_{1}$ as a function of the single modulus $t_{2}$. \ Because a larger amount
of flux passes through the $S^{3}$ corresponding to $t_{1}$, it is enough to
expand $\alpha/N_{1}$ to first order in $t_{1}$ and to leading order in $\log
t_{2}$. \ Extremizing $\alpha/N_{1}$ with respect to $t_{2}$, we find that
$V_{\mathrm{eff}}$ develops a flat direction when:%
\begin{equation}
\frac{8\pi^{2}}{g_{\mathrm{YM}}^{2}\left\vert N_{1}\right\vert }\simeq\frac
{M}{80t_{1}}\text{.} \label{approxbehavioranone}%
\end{equation}
Using equation (\ref{leadstwo}) to approximate the value of $t_{1}$ yields the
value of $\lambda_{1}\equiv g_{\mathrm{YM}}^{2}\left\vert N_{1}\right\vert $
for which the system develops an instability:%
\begin{equation}
\frac{1}{\lambda_{1,\ast}}\simeq-\frac{1}{8\pi^{2}}W_{-1}\left(  -\frac{1}%
{80}\frac{\log\left\vert \frac{\Lambda_{0}}{\Delta}\right\vert ^{2}%
}{\left\vert \frac{\Lambda_{0}}{\Delta}\right\vert ^{2}}\right)
\end{equation}
where $\lambda_{1,\ast}$ denotes the bare 't Hooft coupling at the phase
transition. \ To obtain a crude estimate of when we expect to lose
metastability, we treat the left hand side of equation
(\ref{approxbehavioranone}) as an order one number, obtaining:%
\begin{equation}
\frac{1}{\lambda_{1,\ast}}\sim\frac{1}{8\pi^{2}}\log\left\vert \frac
{\Lambda_{0}}{\Delta}\right\vert ^{2}\text{.}%
\end{equation}
This estimate is in accord with equation (\ref{firstestimate}).

As in subsection \ref{Z2breakdown}, it is important to compute the masses
squared of the bosonic fluctuations at the metastable minimum in order to
determine the mode of instability for this flux configuration. \ In this case,
the appropriate linear combination of fields which diagonalizes the mass
matrix is somewhat messier and we defer the details of this computation to
appendix B. \ We find that the unstable mode corresponds to the smaller branch
cut increasing in size at a much faster rate than its larger counterpart.

\section{Endpoints of a Phase Transition\label{Endpoint}}

For sufficiently large 't Hooft coupling, the metastable vacua present at weak
coupling cease to exist. \ The moduli subsequently roll to larger values so
that a perturbative expansion in the glueball fields $S_{i}$ is no longer
valid. \ As the branch cuts increase in size and meet each other, the 3-cycle
$B_{1}-B_{2}$ reduces to zero size. \ We have checked numerically that over
this range the potential attains a minimal value only once the cuts touch.
\ Because no flux passes through $B_{1}-B_{2}$, the moduli will not be
stabilized away from the corresponding conifold point. \ Near this region, the
contribution of new light states associated to a D3-brane wrapping
$B_{1}-B_{2}$ and an instanton gas of D5-branes all wrapping the same cycle
will determine the low energy dynamics of the theory.

Let us first consider the contribution due to the D5-branes. \ From the
perspective of the spacetime, a D5-brane wrapping $B_{1}-B_{2}$ corresponds to
a domain wall solution which separates vacua with distinct values of the flux.
\ Indeed, for vacua near the semi-classical expansion point, this is the
primary mechanism by which the vacuum can decay to a supersymmetric flux
configuration \cite{ABSV}. \ As $B_{1}-B_{2}$ collapses, the corresponding
action for quantum tunneling tends to zero and a gas of D5-branes will scan
over all available flux configurations. \ This will necessarily change the
shape of the potential for the moduli. \ When enough flux has been
annihilated, it is then possible for the moduli to subsequently either quantum
mechanically tunnel through moduli space or (if the shape of the potential has
changed enough) classically roll back out to the semi-classical regime.
\ Letting the vector $\overrightarrow{N_{\ast}}$ denote the critical value of
the fluxes for which metastability is lost, the flux vector for the new
metastable minimum will differ from $\overrightarrow{N_{\ast}}$ by a finite amount.

Next consider the contribution due to the D3-brane. \ As the 3-cycle
$B_{1}-B_{2}$ collapses, the geometry approaches a conifold point such that
both the K\"{a}hler metric and $V_{\mathrm{eff}}$ develop singularities.
\ Just as in the supersymmetric case, the additional light states which
resolve this singularity correspond to a D3-brane wrapped over $B_{1}-B_{2}$.
\ Because an analysis of a generic flux configuration would require a fairly
precise knowledge of the periods near this region in moduli space, we restrict
our discussion to the case $N_{1}=-N_{2}$ with the branch cuts aligned along
the real axis of the complex $x$-plane. \ In subsection \ref{further} we
present evidence that \textit{if} we ignore the presence of nearly tensionless
domain walls, the light magnetic states of the D3-brane condense and cause
such a flux configuration to transition to a non-K\"{a}hler manifold.

The rather different physical nature of the D5-brane instanton gas and
massless states contributed by the D3-brane indicate that a more detailed
analysis of the endpoint of this phase transition is likely to be quite
difficult. \ Our general expectation, however, is that the contribution due to
the D5-branes will typically cause the moduli to relax back to the
semi-classical expansion point. \ Indeed, in the case $\left\vert
N_{1}\right\vert \gg\left\vert N_{2}\right\vert $, the $N_{2}$ units of flux
can be treated as a probe of the background flux configuration determined by
$N_{1}$. \ In this case it is doubtful that such a small perturbation could
cause the geometry to undergo a transition to a non-K\"{a}hler manifold.

The situation is less clear when $N_{1}\sim-N_{2}$. \ As we show in subsection
\ref{energycutstouch}, when $N_{1}=-N_{2}\equiv N$ and along the locus
$S_{1}=-S_{2}>0$, the value of the potential when the cuts touch is
independent of $N$. \ This implies that while the gas of tensionless domain
walls can still drive the system back to a metastable vacuum, it is not
energetically favorable to eliminate a small amount of flux. \ The additional
light magnetic states due to the D3-brane can then potentially influence the
endpoint of the phase transition.

The rest of this section presents additional details concerning the flux
configuration $N_{1}=-N_{2}\equiv N$ and is organized as follows. \ In
subsection \ref{energycutstouch} we show that when $S_{1}=-S_{2}>0$, the value
of $V_{\mathrm{eff}}$ at the point where the cuts touch is independent of $N$.
\ This computation allows us to determine in subsection \ref{Hysteresis} the
minimal drop in flux necessary to tunnel back out to the semi-classical
regime. \ In subsection \ref{magdescription} we describe in more detail the
singular behavior of $V_{\mathrm{eff}}$ by passing to a new basis of special
coordinates. \ Using this dual magnetic description we show in subsection
\ref{further} that in the absence of D5-brane effects, a condensate of light
magnetic states causes the geometry to transition to a non-K\"{a}hler
manifold.%
\begin{figure}
[ptbh]
\begin{center}
\includegraphics[
height=3.8024in,
width=5.0859in
]%
{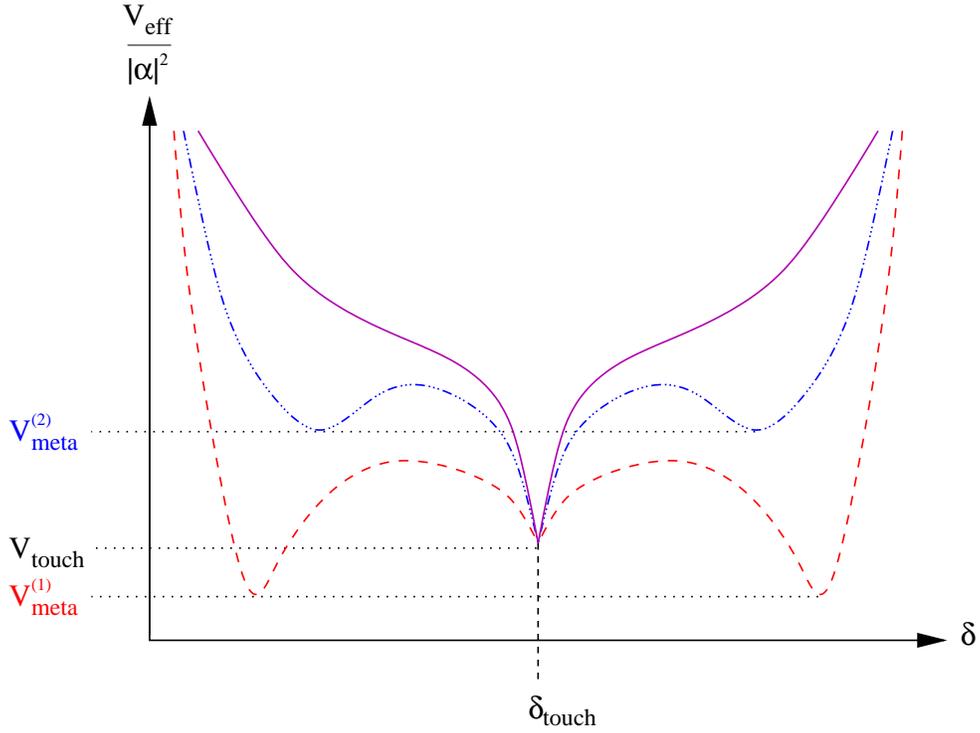}%
\caption{Depiction of $V_{\mathrm{eff}}/\left\vert \alpha\right\vert ^{2}$ as
a function of $\delta\equiv a_{1}^{+}-a_{1}^{-}=a_{2}^{+}-a_{2}^{-}>0$ with
$N_{1}=-N_{2}\equiv N$. \ The value of $V_{\mathrm{eff}}$ at the point where
the cuts touch is independent of $N$. \ When $N$ is small (dashed red), the
system possesses a metastable minimum with energy density $V_{\mathrm{meta}%
}^{(1)}<V_{\mathrm{touch}}$. \ For intermediate values of $N$ which still
admit a metastable minimum (dot dashed blue), the energy density of the vacuum
is $V_{\mathrm{meta}}^{(2)}>V_{\mathrm{touch}}$. \ For large enough values of
$N$ (solid purple), the system undergoes a phase transition and no metastable
minimum exists.}%
\label{hilo}%
\end{center}
\end{figure}

\subsection{Energy Near Cuts Touching\label{energycutstouch}}

We now show that as the cuts touch along the locus $S_{1}=-S_{2}>0$, the
effective potential approaches a constant value independent of $N$. \ By
virtue of equation (\ref{TAUMATrelation}), $V_{\mathrm{eff}}$ may be written
as:%
\begin{equation}
\frac{V_{\mathrm{eff}}}{\left\vert \alpha\right\vert ^{2}}=\frac{1}{\tau
_{++}+2\frac{M-\overline{M}}{2\pi i}}\left(  4i-2i\left(  \det\tau
+\frac{M-\overline{M}}{2\pi i}\tau_{\mathrm{ell}}\right)  \left\vert \frac
{N}{\alpha}\right\vert ^{2}\right)  \label{veffnonetwo}%
\end{equation}
where $\tau_{++}=\tau_{11}+\tau_{22}+2\tau_{12}$ is the \textquotedblleft
center of mass\textquotedblright\ coupling, $\tau_{\mathrm{ell}}=\tau
_{11}+\tau_{22}-2\tau_{12}$ is the complex structure modulus of the elliptic
curve, and as before, $M=\log\left(  \Lambda_{0}^{2}/\Delta^{2}\right)  $.
\ Note that as the cuts touch, $\tau_{\mathrm{ell}}$ tends to zero. \ Because
$\tau_{11}=\tau_{22}$, this implies $\tau_{12}$ tends to $\tau_{11}$ and thus
$\det\tau$ tends to zero. $\ $We therefore conclude that $V_{\mathrm{eff}}$
approaches the value:%
\begin{equation}
V_{\mathrm{eff}}\rightarrow\frac{4i\left\vert \alpha\right\vert ^{2}}%
{4\tau_{11}+2\frac{M-\overline{M}}{2\pi i}}\text{.}%
\end{equation}
Next, recall from appendix A that the behavior of $\tau_{11}$ is of the form
$(B-M)/2\pi i$ where $B$ is a function of the moduli which is independent of
$\alpha,N$ and $M$. \ The value of $V_{\mathrm{touch}}$ is then:%
\begin{equation}
V_{\mathrm{touch}}=\frac{2\pi\left\vert \alpha\right\vert ^{2}}{B_{\mathrm{t}%
}+\log\left\vert \frac{\Lambda_{0}}{\Delta}\right\vert ^{2}} \label{VTOUCH}%
\end{equation}
where $B_{\mathrm{t}}$ denotes the value of $B$ when the cuts touch. \ This is
manifestly independent of $N$. \ Finally, we note that as $\Delta$ tends to
zero, so too does $V_{\mathrm{touch}}$.

\subsection{Flux Hysteresis\label{Hysteresis}}

Close to the region in moduli space where the cuts touch, a gas of nearly
tensionless D5-branes can cause the system to tunnel back out to a metastable
configuration with lower flux. \ This tunneling will cause the flux to undergo
hysteresis. \ In this subsection we determine the minimal size of this jump in
the flux number. \ Turning the discussion around, this also determines the
range of values for which a metastable vacuum near the semi-classical
expansion point can tunnel to the region where the cuts nearly touch.%
\begin{figure}
[ptbh]
\begin{center}
\includegraphics[
height=3.8024in,
width=5.0859in
]%
{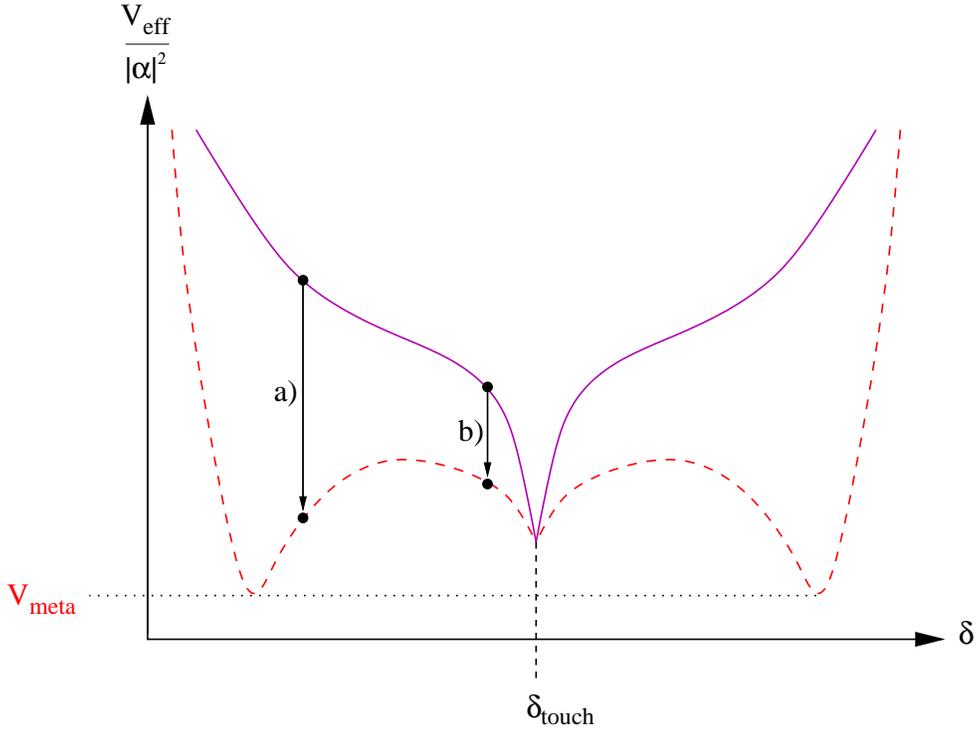}%
\caption{Depiction of $V_{\mathrm{eff}}/\left\vert \alpha\right\vert ^{2}$ as
a function of $\delta\equiv a_{1}^{+}-a_{1}^{-}=a_{2}^{+}-a_{2}^{-}>0$ with
$N_{1}=-N_{2}\equiv N$. \ The figure shows flux configurations which possess a
metastable minimum (dashed red) and those which do not (solid purple). \ Close
to the region in moduli space where the cycle $B_{1}-B_{2}$ collapses, the
presence of nearly tensionless domain walls can cause the amount of flux to
jump. \ To reach a metastable minimum near the semi-classical expansion point,
the moduli then either classically roll (a) or tunnel through moduli space
(b).}%
\label{fluxjump}%
\end{center}
\end{figure}

We begin by characterizing the possible transition points for tunneling
processes along the locus $S_{1}=-S_{2}>0$. \ In addition to the metastable
vacua near the semi-classical expansion point, there is another vacuum with
\textit{identical} energy where the branch cuts overlap almost completely.
\ To increase the scope of our discussion, we briefly consider more general
configurations such that $S_{1}>0>S_{2}$. \ With notation as in section
\ref{twocut}, the endpoints of the branch cuts in the semi-classical regime
satisfy $a_{1}^{+}>a_{1}^{-}>a_{2}^{+}>a_{2}^{-}$\ with\ special coordinates
defined by the integrals:%
\begin{equation}
S_{1}=\frac{1}{2\pi i}\underset{a_{1}^{-}}{\overset{a_{1}^{+}}{\int}}ydx\text{
\ \ \ \ }S_{2}=-\frac{1}{2\pi i}\underset{a_{2}^{-}}{\overset{a_{2}^{+}}{\int
}}ydx\text{.}%
\end{equation}
where the choice of signs is dictated by branch cut considerations. \ In the
region where the branch cuts overlap, we instead have $a_{1}^{+}>a_{2}%
^{+}>a_{1}^{-}>a_{2}^{-}$ with special coordinates defined by the integrals:%
\begin{equation}
S_{1}^{\prime}=\frac{1}{2\pi i}\underset{a_{2}^{+}}{\overset{a_{1}^{+}}{\int}%
}ydx\ \ \ \ S_{2}^{\prime}=-\frac{1}{2\pi i}\underset{a_{2}^{-}}%
{\overset{a_{1}^{-}}{\int}}ydx.
\end{equation}
It therefore follows from geometric considerations that there is a highly
non-trivial duality in the low energy effective field theory between vacua
with $S_{i}$ small and vacua with $S_{i}^{\prime}$ small. \ In addition to
these two physically indistinguishable configurations, there is the local
singular minimum where the cuts touch.

The system can only tunnel from the region where the cuts touch to a
metastable vacuum with small $S_{i}$ when:%
\begin{equation}
V_{\mathrm{meta}}\leq V_{\mathrm{touch}} \label{inequtouchmeta}%
\end{equation}
where $V_{\mathrm{meta}}$ denotes the value of $V_{\mathrm{eff}}$ evaluated at
such a metastable critical point and $V_{\mathrm{touch}}$ is given by equation
(\ref{VTOUCH}). \ Approximating $V_{\mathrm{meta}}$ by equation
(\ref{energyleading}), the flux must therefore satisfy the bound:%
\begin{equation}
g_{\mathrm{YM}}^{2}\left\vert N\right\vert \geq\frac{4\pi^{2}}{\log\left\vert
\frac{\Lambda_{0}}{\Delta}\right\vert ^{2}}\left(  1-\sqrt{\frac
{B_{\mathrm{t}}}{B_{\mathrm{t}}+\log\left\vert \frac{\Lambda_{0}}{\Delta
}\right\vert ^{2}}}\right)  \text{.} \label{bounder}%
\end{equation}
It now follows from equations (\ref{breakdown}) and (\ref{bounder}) that in
jumping from a flux configuration which does not admit a metastable vacuum
near the semi-classical expansion point to one which does, the flux drops by
an amount:%
\begin{equation}
\delta\left\vert N\right\vert \geq\frac{4\pi^{2}}{g_{\mathrm{YM}}^{2}%
\log\left\vert \frac{\Lambda_{0}}{\Delta}\right\vert ^{2}}\sqrt{\frac
{B_{\mathrm{t}}}{B_{\mathrm{t}}+\log\left\vert \frac{\Lambda_{0}}{\Delta
}\right\vert ^{2}}}\text{.}%
\end{equation}

Turning the discussion around, the range of fluxes for which it is possible to
tunnel from a metastable vacuum near the semi-classical expansion point to the
region where the cuts touch is:%
\begin{equation}
\frac{4\pi^{2}}{\log\left\vert \frac{\Lambda_{0}}{\Delta}\right\vert ^{2}%
}\gtrsim g_{\mathrm{YM}}^{2}\left\vert N\right\vert \geq\frac{4\pi^{2}}%
{\log\left\vert \frac{\Lambda_{0}}{\Delta}\right\vert ^{2}}\left(
1-\sqrt{\frac{B_{\mathrm{t}}}{B_{\mathrm{t}}+\log\left\vert \frac{\Lambda_{0}%
}{\Delta}\right\vert ^{2}}}\right)  \label{window}%
\end{equation}
where the crude upper bound follows from equation (\ref{breakdown}) and the
requirement that a metastable minimum exists. \ Note in particular that the
admissible range of values of the 't Hooft coupling which permit such a
process is parametrically tied to the value for which metastability is lost.
\ Thus, nearly as soon as we increase the 't Hooft coupling to a value where
this quantum effect can contribute, it is overtaken by classical effects.

\subsection{Dual Magnetic Description\label{magdescription}}

Near the region in moduli space where the cuts touch, the fields $S_{1}$ and
$S_{2}$ have become large and it is appropriate to change to a dual magnetic
basis of fields. \ From the perspective of the geometry, this corresponds to
performing a change of basis which preserves the intersection pairing of the
geometry. \ Using this dual basis, we now show that the effective potential
develops a cusp when the cuts touch\footnote{Recall that a differentiable
function $f(x)$ of a single real variable is said to have a cusp at the point
$a$ if $f^{\prime}(x)\rightarrow\pm\infty$ as $x\rightarrow a^{\mp}$. \ A
similar definition holds for functions of several variables.}.

The appropriate change of basis is dictated by the geometry of the Riemann
surface. \ Along the entire locus considered, $S_{1}+S_{2}$ has remained zero.
\ Further, the cycle $B_{1}-B_{2}$ is close to zero size. \ This implies that
the new $A$-cycles are $\widetilde{A}_{1}=B_{2}-B_{1}$ and $\widetilde{A}%
_{2}=A_{1}+A_{2}$. \ Dual to these are new $B$-cycles $\widetilde{B}_{1}%
=A_{1}$ and $\widetilde{B}_{2}=B_{2}$. \ Note that this new basis preserves
the intersection pairing of the geometry. \ The dual magnetic coordinates
$\widetilde{S}_{i}$ and $\widetilde{\Pi}_{i}$ are therefore related to the
original special coordinates by:%
\begin{align}
\widetilde{S}_{1}  &  =\Pi_{2}-\Pi_{1}\text{ \ \ \ \ \ \ \ \ }\widetilde
{S}_{2}=S_{1}+S_{2}\\
\widetilde{\Pi}_{1}  &  =S_{1}\text{ \ \ \ \ \ \ \ \ \ \ \ \ \ \ \ }%
\widetilde{\Pi}_{2}=\Pi_{2}\text{. \ \ \ \ }%
\end{align}
It follows from Picard-Lefschetz singularity theory that the only non-trivial
monodromy arises from the transformation $\widetilde{S}_{1}\mapsto e^{2\pi
i}\widetilde{S}_{1}$. \ We therefore conclude that the leading order behavior
of $\widetilde{\Pi}_{2}$ is regular and $\widetilde{\Pi}_{1}$ depends
logarithmically on $\widetilde{S}_{1}$:%
\begin{equation}
\widetilde{\Pi}_{1}=\frac{1}{2\pi i}\widetilde{S}_{1}\log\frac{\widetilde
{S}_{1}}{g\Delta^{3}}+O(\widetilde{S}_{1}^{0})\text{.}%
\end{equation}
For a general flux configuration, the induced superpotential in the new
coordinates is:%
\begin{equation}
\mathcal{W}_{\mathrm{eff}}=\int H_{3}\wedge\Omega=\alpha\widetilde{S}%
_{2}+\left(  N_{+}+N_{-}\right)  \widetilde{S}_{1}+2N_{+}\widetilde{\Pi}_{2}
\label{magsusypot}%
\end{equation}
where $N_{\pm}\equiv\left(  N_{2}\pm N_{1}\right)  /2$. \ Note that
$\mathcal{W}_{\mathrm{eff}}$ is independent of $\widetilde{\Pi}_{1}$. \ In
particular, when $N_{1}=-N_{2}\equiv N$, the superpotential is independent of
both of the $\widetilde{\Pi}_{i}$'s. \ The entries of the new period matrix
are:%
\begin{equation}
\widetilde{\tau}_{ij}=\frac{\partial\widetilde{\Pi}_{j}}{\partial\widetilde
{S}_{i}}.
\end{equation}
Because $\widetilde{\Pi}_{2}$ is regular, we may approximate the K\"{a}hler
metric as:%
\begin{equation}
\operatorname{Im}\widetilde{\tau}_{ij}=-\frac{1}{4\pi}\left[
\begin{array}
[c]{cc}%
\log\left\vert \frac{\widetilde{S}_{1}}{g\Delta^{3}}\right\vert ^{2} &
c_{12}\\
c_{12} & c_{22}%
\end{array}
\right]
\end{equation}
where the $c_{ij}$ are non-zero constants. \ In the new coordinates, the
effective potential is:%
\begin{equation}
V_{\mathrm{eff}}\left(  \widetilde{S}_{1},\widetilde{S}_{2}\right)  =\left[
\begin{array}
[c]{cc}%
-N & \alpha
\end{array}
\right]  \frac{1}{\operatorname{Im}\widetilde{\tau}}\left[
\begin{array}
[c]{c}%
-N\\
\overline{\alpha}%
\end{array}
\right]
\end{equation}
in the obvious notation. \ It therefore follows that $V_{\mathrm{eff}}$
attains a minimum at $\widetilde{S}_{1}=0$.

Replacing the $\widetilde{S}_{i}$ by fields $\widetilde{X}_{i}$ with
canonically normalized kinetic terms\footnote{Although such a field
redefinition will in general introduce anomalies into the Lagrangian density,
this will not alter the conclusions of our analysis.}, the derivative of
$V_{\mathrm{eff}}$ with respect to $\widetilde{X}_{1}$ is:%
\begin{equation}
\frac{\partial V_{\mathrm{eff}}}{\partial\widetilde{X}_{1}}\sim\frac
{1}{\widetilde{S}_{1}}\frac{1}{f(\widetilde{S}_{1},\widetilde{S}_{2})}%
\end{equation}
where $f$ contains at most $\log\widetilde{S}_{1}$ type divergences. \ We
therefore conclude that the potential has a cusp at $\widetilde{S}_{1}=0$.

\subsection{A Further Phase Transition\label{further}}

The singular behavior of the effective potential implies the presence of
additional light states which have been integrated out. \ As in the case of
the conifold \cite{StromingerConifold,GreeneMorrisonStromingerConifold}, these
light degrees of freedom correspond to a D3-brane wrapping the vanishing
3-cycle $B_{1}-B_{2}$. \ In $\mathcal{N}=2$ language, this corresponds to a
hypermultiplet which is charged under the gauge boson of the $U(1)$ vector
multiplet with scalar component $\widetilde{S}_{1}$. \ The superpotential is
now:%
\begin{equation}
\mathcal{W}_{\mathrm{eff}}=\alpha\widetilde{S}_{2}+\left(  N_{+}+N_{-}\right)
\widetilde{S}_{1}+2N_{+}\widetilde{\Pi}_{2}+h\widetilde{S}_{1}Q_{L}Q_{R}%
\end{equation}
where $h$ is a Yukawa coupling and $Q_{L}$ and $Q_{R}$ denote the
$\mathcal{N}=1$ chiral multiplets of the hypermultiplet. \ At a critical point
of $V_{\mathrm{eff}}$, the $Q$'s condense with expectation value:%
\begin{equation}
\left\langle Q_{L}Q_{R}\right\rangle =\frac{N}{h}\text{.}%
\end{equation}
This condensate signals the presence of a new holomorphic 2-cycle
$\Sigma=\partial A_{1}$ in the geometry with size $\left\vert N/h\right\vert
$. \ Perhaps surprisingly, the resulting manifold is non-K\"{a}hler!

To see this, let us suppose to the contrary that the new geometry is a
Calabi-Yau threefold. \ In this case, the absence of any other normalizable
$(1,1)$ forms implies that the $(1,1)$ form $k_{\Sigma}$ which measures the
volume of $\Sigma$ determines the local metric of the new geometry. \ It now
follows from Stokes' theorem that:%
\begin{equation}
\underset{A_{1}}{\int}dk_{\Sigma}=\underset{\Sigma}{\int}k_{\Sigma}%
\neq0\text{.}%
\end{equation}
This implies that $k_{\Sigma}$ is not closed. \ We therefore conclude that the
resulting manifold is not a Calabi-Yau threefold, but instead belongs to the
category of \textit{generalized} Calabi-Yau threefolds
\cite{HitchinGeneralized}. \ While it is doubtful that the resulting physical
configuration is supersymmetric or even metastable, a proper analysis is
beyond the scope of this paper and we defer a full study of this question to
future work.

\section{Conclusions\label{Conclusions}}

In this paper we have studied the phase structure of a strongly coupled
supersymmetry breaking configuration of D5-branes and anti-D5-branes wrapped
over homologous rigid $S^{2}$'s of a non-compact Calabi-Yau threefold using
the large $N$ dual description of this system. \ In much of this paper we
focused on the closed string dual geometry with two branch cuts. \ Even in
this simple case, higher order corrections to the potential for the glueball
fields generate an elaborate phase structure which can already be seen at the
two loop level. \ Near the semi-classical expansion point, this two loop
effect lifts the degeneracy in energy density between the many confining vacua
of the theory. \ When the scale of confinement is not exponentially
suppressed, this generates a large number of additional metastable vacua.
\ For sufficiently large values of the 't Hooft coupling this same effect also
lifts the metastable vacua present at weak coupling. \ After this phase
transition, the branch cuts expand in size until they are close to touching.
\ Although the presence of nearly tensionless domain walls close to this
region of moduli space will most likely cause the system to relax to a
metastable vacuum of lower flux, the presence of new massless states when the
cuts touch may also allow the geometry to transition to a non-K\"{a}hler
manifold. \ We now discuss some implications of this work.

As the separation $\Delta$ between the branes decreases, the glueball
potential develops an instability. \ Although it is tempting to identify this
instability with the open string theory tachyon, there is a potentially
serious problem with this interpretation. \ Indeed, it follows from equation
(\ref{breakdown}) that the effective potential develops an instability when:%
\begin{equation}
\frac{1}{g_{\mathrm{YM}}^{2}\left\vert N\right\vert }\sim\log\left\vert
\frac{\Lambda_{0}}{\Delta}\right\vert ^{2}\text{.}%
\end{equation}
On the other hand, the tachyonic mode of the brane/anti-brane system is
independent of $N$ because this constant factors out of all relevant open
string amplitudes. \ We therefore conclude that an identification of the two
instabilities is not naively correct.

We have also seen a preliminary indication that the number of critical points
of $V_{\mathrm{eff}}$ crucially depends on the amount of flux in the closed
string holographic dual. \ As this amount of flux changes, a local maximum and
minimum may merge. \ Such a change in the number of metastable minima cannot
be detected by a Morse-theoretic index. \ This may have implications for
recent attempts to count the number of supersymmetry breaking vacua in flux
compactifications. \ Indeed, many of the techniques developed thus far rely on
similar indices to count the number of admissible vacua \cite{AshokDouglas}.
\ It would therefore be interesting to determine whether such methods properly
account for the metastable vacua studied in this paper.

Although a more detailed analysis of the phase structure near the region of
moduli space where the branch cuts touch will most likely be difficult, we
have seen that when $N_{1}\sim-N_{2}$, the geometry may undergo a further
phase transition to a non-K\"{a}hler manifold. \ Even so, the decay of the
vacuum due to flux line annihilation may obstruct this intriguing possibility
from contributing to the phase structure of the theory. \ It is likely that a
proper description of the effective theory near this region in moduli space
will require a more generalized effective potential which treats both the
moduli and the fluxes as dynamical variables. \ \textit{Even if} the geometry
can transition to a non-K\"{a}hler manifold, the resulting configuration is
unlikely to be stable. \ It would be interesting to determine the endpoint of
this further phase transition.

In much of this paper we restricted our analysis to the two cut geometry.
\ While this should provide an adequate characterization of \textquotedblleft
two body\textquotedblright\ interactions, it is possible that the interaction
of three or more cuts could lead to further novel phases. \ Although we still
expect the phases of the glueball fields to align in an energetically
preferred configuration, the relative orientation between the cuts will depend
on the location of the $a_{i}$. \ Indeed, treating the branch cuts as small
dipole moments, an energetically preferred configuration may be frustrated for
a large lattice of cuts, much as in the two dimensional Ising model on a
triangular lattice. \ It is well-known in the setting of condensed matter
systems that magnetic frustration can produce novel phases such as spin
liquids and glasses. \ It is therefore likely that a similarly rich class of
phenomena are present in metastable multi-cut geometries.

\section*{Acknowledgements}

We thank M. Aganagic, N. Arkani-Hamed, M. Huang, A. Klemm, J. Lapan, and X.
Yin for helpful discussions. \ CV thanks the CTP\ at MIT for hospitality
during his sabbatical leave. \ The work of the authors is supported in part by
NSF grants PHY-0244821 and DMS-0244464. \ The research of JJH\ is also
supported by an NSF\ Graduate Fellowship, and the research of JS is also
supported by the Korea Foundation for Advanced Studies.

\appendix

\section*{Appendix A: Two Cut Semi-Classical $\tau_{ij}$ and $\mathcal{F}%
_{ijk}$}

In this appendix we collect explicit expressions for $\tau_{ij}$ and
$\mathcal{F}_{ijk}=\partial_{i}\tau_{jk}$ for the two cut geometry defined by
equation (\ref{defcon}). \ Setting $\Delta\equiv a_{1}-a_{2}$, $t_{1}\equiv
S_{1}/g\Delta^{3}$ and $t_{2}\equiv-S_{2}/g\Delta^{3}$, we have \cite{CIV}:%
\begin{align}
2\pi i\tau_{11}  &  =\log t_{1}-\log\frac{\Lambda_{0}^{2}}{\Delta^{2}}+\left(
4t_{1}+10t_{2}\right)  +\left(  32t_{1}^{2}+182t_{1}t_{2}+118t_{2}^{2}\right)
+O(t^{3})\label{TAUONEONE}\\
2\pi i\tau_{12}  &  =-\log\frac{\Lambda_{0}^{2}}{\Delta^{2}}+\left(
-10t_{1}-10t_{2}\right)  +\left(  -91t_{1}^{2}-236t_{1}t_{2}-91t_{2}%
^{2}\right)  +O(t^{3})\label{TAUONETWO}\\
2\pi i\tau_{22}  &  =\log t_{2}-\log\frac{\Lambda_{0}^{2}}{\Delta^{2}}+\left(
4t_{2}+10t_{1}\right)  +\left(  32t_{2}^{2}+182t_{1}t_{2}+118t_{1}^{2}\right)
+O(t^{3}), \label{TAUTWOTWO}%
\end{align}
and:%
\begin{align}
2\pi ig\Delta^{3}\mathcal{F}_{111}  &  =\frac{1}{t_{1}}+4+\left(
64t_{1}+182t_{2}\right)  +O(t^{2})\label{FOOO}\\
2\pi ig\Delta^{3}\mathcal{F}_{112}  &  =-10+\left(  -182t_{1}-236t_{2}\right)
+O(t^{2})\\
2\pi ig\Delta^{3}\mathcal{F}_{122}  &  =10+236t_{1}+182t_{2}+O(t^{2})\\
2\pi ig\Delta^{3}\mathcal{F}_{222}  &  =-\frac{1}{t_{2}}-4-64t_{2}%
-182t_{1}+O(t^{2}). \label{FTTT}%
\end{align}

\section*{Appendix B: Mass Spectrum for $\left\vert N_{1}\right\vert
\gg\left\vert N_{2}\right\vert $}

In this appendix we compute the bosonic and fermionic masses for flux
configurations with $\left\vert N_{1}\right\vert \gg\left\vert N_{2}%
\right\vert $, $\theta_{\mathrm{YM}}=0$ and $S_{1}>0>S_{2}$. \ With the
kinetic terms of the Lagrangian density canonically normalized, the $4\times4$
bosonic mass squared matrix $m_{\mathrm{Bosonic}}^{2}$ takes the block
diagonal form:%
\begin{equation}
m_{\mathrm{Bosonic}}^{2}=A^{(R)}\oplus A^{(I)}%
\end{equation}
where the $A^{(R,I)}$ are $2\times2$ matrices of the form:%
\begin{equation}
A^{(R,I)}=\left(
\begin{array}
[c]{cc}%
\frac{\left(  \partial_{1}^{(R,I)}+\partial_{2}^{(R,I)}\right)  ^{2}%
V_{\mathrm{eff}}}{1+v} & -\frac{\left(  \partial_{1}^{(R,I)}-\partial
_{2}^{(R,I)}\right)  \left(  \partial_{1}^{(R,I)}+\partial_{2}^{(R,I)}\right)
V_{\mathrm{eff}}}{\sqrt{1-v^{2}}}\\
-\frac{\left(  \partial_{1}^{(R,I)}-\partial_{2}^{(R,I)}\right)  \left(
\partial_{1}^{(R,I)}+\partial_{2}^{(R,I)}\right)  V_{\mathrm{eff}}}%
{\sqrt{1-v^{2}}} & \frac{\left(  \partial_{1}^{(R,I)}-\partial_{2}%
^{(R,I)}\right)  ^{2}V_{\mathrm{eff}}}{1-v}%
\end{array}
\right)
\end{equation}
and $A^{(R)}$ ($A^{(I)}$) corresponds to the mass matrix for the real
(imaginary) components of the $S_{i}$'s. \ In the above we have defined:%
\begin{equation}
\partial_{j}^{(R)}=\frac{1}{\sqrt{\operatorname{Im}\tau_{jj}}}\frac{\partial
}{\partial\operatorname{Re}S_{j}},\text{ }\partial_{j}^{(I)}=\frac{1}%
{\sqrt{\operatorname{Im}\tau_{jj}}}\frac{\partial}{\partial\operatorname{Im}%
S_{j}},\text{ }v=\frac{\operatorname{Im}{{\tau}_{12}}^{2}}%
{\mathrm{\operatorname{Im}}{\tau}_{11}\operatorname{Im}{\tau}_{22}},
\end{equation}
and for future use we also introduce:%
\begin{equation}
a=\frac{|N_{1}|}{2\pi t_{1}\operatorname{Im}{\tau}_{11}}\text{, }%
b=\frac{|N_{2}|}{2\pi t_{2}\operatorname{Im}{\tau}_{22}}.
\end{equation}
In the above expressions the components of $\tau_{ij}$ correspond to their
value at the critical point of $V_{\mathrm{eff}}$ and should therefore be
treated as constants. \ When $\left\vert N_{1}\right\vert >>|N_{2}|$, the
masses squared and eigenmodes of the block $A^{(R)}$ are:%
\begin{align}
m_{\operatorname{Re}S_{1}}^{2}  &  =\frac{b^{2}}{(1-v)^{2}} & \left(
\sqrt{\frac{1-\sqrt{v}}{1+\sqrt{v}}},1\right)  _{R}  &  \oplus\left(
0,0\right)  _{I}\\
m_{\operatorname{Re}S_{2}}^{2}  &  =a^{2}-2a\left(  10|N_{2}|-\frac
{2|N_{1}|+5|N_{2}|}{\operatorname{Im}{\tau}_{11}\pi}\right)  & \left(
-\sqrt{\frac{1+\sqrt{v}}{1-\sqrt{v}}},1\right)  _{R}  &  \oplus\left(
0,0\right)  _{I}%
\end{align}
and the masses squared and eigenmodes of the block $A^{(I)}$ are similarly:%
\begin{align}
m_{\operatorname{Im}S_{1}}^{2}  &  =\frac{b^{2}}{(1-v)^{2}} & \left(
\sqrt{\frac{1-\sqrt{v}}{1+\sqrt{v}}},1\right)  _{I}\oplus &  \left(
0,0\right)  _{R}\\
m_{\operatorname{Im}S_{2}}^{2}  &  =a^{2}+2a\left(  10|N_{2}|+\frac
{2|N_{1}|+5|N_{2}|}{\operatorname{Im}{\tau}_{11}\pi}\right)  & \left(
-\sqrt{\frac{1+\sqrt{v}}{1-\sqrt{v}}},1\right)  _{I}\oplus &  \left(
0,0\right)  _{R}\text{.}%
\end{align}
Grouping the fermions according to the supermultiplet structure inherited from
the $\mathcal{N}=1$ supersymmetry of the branes, the non-zero fermion masses
are:%
\begin{align}
m_{\psi_{S}}  &  =\frac{1}{1-v}\left(  a+\frac{2|N_{1}|+5|N_{2}|+10|N_{1}%
|\frac{\operatorname{Im}{\tau}_{12}}{\operatorname{Im}{\tau}_{22}}%
}{\operatorname{Im}{\tau}_{11}\pi}\right) \\
m_{\psi_{A}}  &  =\frac{1}{1-v}\left(  b+\frac{2|N_{2}|+5|N_{1}|+10|N_{1}%
|\frac{\operatorname{Im}{\tau}_{12}}{\operatorname{Im}{\tau}_{11}}%
}{\operatorname{Im}{\tau}_{22}\pi}\right)
\end{align}
with similar notation to that given above equation (\ref{fermionmass}). \ By
inspection of the above formulae, we see that the two loop correction
increases the difference between the bosonic and fermionic masses already
present at leading order.

Keeping $g_{\mathrm{YM}}^{2}\left\vert N_{2}\right\vert $ fixed, we now
determine the mode which develops an instability as the 't Hooft coupling
$g_{\mathrm{YM}}^{2}\left\vert N_{1}\right\vert $ approaches the critical
value where the original metastable vacua disappear. \ The determinant of each
block of the mass matrix is:%
\begin{align}
\det A^{(R)}  &  =\frac{4096\pi^{8}\log t_{1}\log t_{2}}{g_{YM}^{8}t_{1}%
^{2}t_{2}^{2}\left(  \log t_{1}\log t_{2}-\log\frac{|{\Lambda}_{0}|^{2}%
}{{|\Delta|}^{2}}\left(  \log t_{1}+\log t_{2}\right)  \right)  ^{6}}\\
&  \times\left(
\begin{array}
[c]{l}%
\log t_{1}\log t_{2}-20t_{1}\left(  \log t_{1}\right)  ^{2}\left(  \log
\frac{|{\Lambda}_{0}|^{2}}{{|\Delta|}^{2}}-\log t_{1}\right) \\
-20t_{2}\left(  \log t_{2}\right)  ^{2}\left(  \log\frac{|{\Lambda}_{0}|^{2}%
}{{|\Delta|}^{2}}-\log t_{2}\right)  +\cdots
\end{array}
\right) \\
\det A^{(I)}  &  =\frac{4096\pi^{8}\log t_{1}\log t_{2}}{g_{YM}^{8}t_{1}%
^{2}t_{2}^{2}\left(  \log t_{1}\log t_{2}-\log\frac{|{\Lambda}_{0}|^{2}%
}{{|\Delta|}^{2}}\left(  \log t_{1}+\log t_{2}\right)  \right)  ^{6}}\\
&  \times\left(
\begin{array}
[c]{l}%
\log t_{1}\log t_{2}+20t_{1}\left(  \log t_{1}\right)  ^{2}\left(  \log
\frac{|{\Lambda}_{0}|^{2}}{{|\Delta|}^{2}}-\log t_{1}\right) \\
+20t_{2}\left(  \log t_{2}\right)  ^{2}\left(  \log\frac{|{\Lambda}_{0}|^{2}%
}{{|\Delta|}^{2}}-\log t_{2}\right)  +\cdots
\end{array}
\right)  .
\end{align}
It follows from the last line of each expression that only
$m_{\operatorname{Re}S_{1}}^{2}$ or $m_{\operatorname{Re}S_{2}}^{2}$ can
vanish. \ Furthermore, because $m_{\operatorname{Re}S_{1}}^{2}%
>>m_{\operatorname{Re}S_{2}}^{2}$, the mode of instability will cause the
smaller cut to expand towards the larger cut. \ For $\left\vert N_{1}%
\right\vert \gg\left\vert N_{2}\right\vert ~$this occurs at a value of $t_{1}$
given by:%
\begin{equation}
1\sim20\left(  -\log t_{1}+\log\left\vert \frac{{\Lambda}_{0}}{{\Delta}%
}\right\vert ^{2}\right)  t_{1}.
\end{equation}
Note that this is similar in form to the critical value of the moduli found in
equation (\ref{rstar}) for the case $N_{1}=-N_{2}$.\newpage
\bibliographystyle{ssg}
\bibliography{phases}

\begingroup\raggedright\begin{thebibliography}{10}

\bibitem{VafaLargeN}
C.~Vafa, ``Superstrings and Topological Strings at Large N,'' {\em J. Math.
  Phys.} {\bf 42} (2001) 2798--2817,
  \href{http://xxx.lanl.gov/abs/hep-th/0008142}{{\tt hep-th/0008142}}.

\bibitem{KachruPearsonVerlinde}
S.~Kachru, J.~Pearson, and H.~Verlinde, ``Brane/Flux Annihilation and the
  String Dual of a Non-Supersymmetric Field Theory,'' {\em JHEP} {\bf 0206}
  (2002) 021, \href{http://xxx.lanl.gov/abs/hep-th/0112197}{{\tt
  hep-th/0112197}}.

\bibitem{KKLT}
S.~Kachru, R.~Kallosh, A.~Linde, and S.~P. Trivedi, ``de Sitter Vacua in String
  Theory,'' {\em Phys. Rev. D} {\bf 68} (2003) 046005,
  \href{http://xxx.lanl.gov/abs/hep-th/0301240}{{\tt hep-th/0301240}}.

\bibitem{KachruMcGreevy}
S.~Kachru and J.~McGreevy, ``Supersymmetric Three-cycles and (Super)symmetry
  Breaking,'' {\em Phys. Rev. D} {\bf 61} (2000) 026001,
  \href{http://xxx.lanl.gov/abs/hep-th/9908135}{{\tt hep-th/9908135}}.

\bibitem{KachruFranco}
R.~Argurio, M.~Bertolini, S.~Franco, and S.~Kachru, ``Gauge/gravity duality and
  meta-stable dynamical supersymmetry breaking,'' {\em JHEP} {\bf 0701} (2007)
  083, \href{http://xxx.lanl.gov/abs/hep-th/0610212}{{\tt hep-th/0610212}}.

\bibitem{ABSV}
M.~Aganagic, C.~Beem, J.~Seo, and C.~Vafa, ``Geometrically Induced
  Metastability and Holography,''
  \href{http://xxx.lanl.gov/abs/hep-th/0610249}{{\tt hep-th/0610249}}.

\bibitem{VerlindeMonopole}
H.~Verlinde, ``On Metastable Branes and a New Type of Magnetic Monopole,''
  \href{http://xxx.lanl.gov/abs/hep-th/0611069}{{\tt hep-th/0611069}}.

\bibitem{ISS}
K.~Intriligator, N.~Seiberg, and D.~Shih, ``Dynamical SUSY Breaking in
  Meta-Stable Vacua,'' {\em JHEP} {\bf 0604} (2006) 021,
  \href{http://xxx.lanl.gov/abs/hep-th/0602239}{{\tt hep-th/0602239}}.

\bibitem{OoguriOokouchi}
H.~Ooguri and Y.~Ookouchi, ``Meta-Stable Supersymmetry Breaking Vacua on
  Intersecting Branes,'' {\em Phys. Lett. B} {\bf 641} (2006) 323--328,
  \href{http://xxx.lanl.gov/abs/hep-th/0607183}{{\tt hep-th/0607183}}.

\bibitem{FrancoUranga}
S.~Franco, I.~Garcia-Etxebarria, and A.~M. Uranga, ``Non-supersymmetric
  Meta-stable Vacua from Brane Configurations,'' {\em JHEP} {\bf 0701} (2007)
  085, \href{http://xxx.lanl.gov/abs/hep-th/0607218}{{\tt hep-th/0607218}}.

\bibitem{MQCDSeibergShih}
I.~Bena, E.~Gorbatov, S.~Hellerman, N.~Seiberg, and D.~Shih, ``A Note on
  (Meta)stable Brane Configurations in MQCD,'' {\em JHEP} {\bf 0611} (2006)
  088, \href{http://xxx.lanl.gov/abs/hep-th/0608157}{{\tt hep-th/0608157}}.

\bibitem{TatarMeta}
R.~Tatar and B.~Wetenhall, ``Metastable Vacua, Geometrical Engineering and MQCD
  Transitions,'' {\em JHEP} {\bf 0702} (2007) 020,
  \href{http://xxx.lanl.gov/abs/hep-th/0611303}{{\tt hep-th/0611303}}.

\bibitem{juanAdS}
J.~M. Maldacena, ``The Large N Limit of Superconformal field theories and
  supergravity,'' {\em Adv. Theor. Math. Phys.} {\bf 2} (1998) 231--252,
  \href{http://xxx.lanl.gov/abs/hep-th/9711200}{{\tt hep-th/9711200}}.

\bibitem{gkPol}
S.~S. Gubser, I.~R. Klebanov, and A.~M. Polyakov, ``Gauge Theory Correlators
  from Non-Critical String Theory,'' {\em Phys. Lett.} {\bf B428} (1998)
  105--114, \href{http://xxx.lanl.gov/abs/hep-th/9802109}{{\tt
  hep-th/9802109}}.

\bibitem{witHolOne}
E.~Witten, ``Anti De Sitter Space and Holography,'' {\em Adv. Theor. Math.
  Phys.} {\bf 2} (1998) 253--291,
  \href{http://xxx.lanl.gov/abs/hep-th/9802150}{{\tt hep-th/9802150}}.

\bibitem{KlebanovStrassler}
I.~R. Klebanov and M.~J. Strassler, ``Supergravity and a Confining Gauge
  Theory: Duality Cascades and $\chi$SB-Resolution of Naked Singularities,''
  {\em JHEP} {\bf 0008} (2000) 052,
  \href{http://xxx.lanl.gov/abs/hep-th/0007191}{{\tt hep-th/0007191}}.

\bibitem{MaldacenaNunezYangMills}
J.~M. Maldacena and C.~Nunez, ``Towards the large $N$ limit of pure
  $\mathcal{N}=1$ Super Yang Mills,'' {\em Phys. Rev. Lett.} {\bf 86} (2001)
  588--591, \href{http://xxx.lanl.gov/abs/hep-th/0008001}{{\tt
  hep-th/0008001}}.

\bibitem{CIV}
F.~Cachazo, K.~Intriligator, and C.~Vafa, ``A Large N Duality via a Geometric
  Transition,'' {\em Nucl. Phys. B} {\bf 603} (2001) 3--41,
  \href{http://xxx.lanl.gov/abs/hep-th/0103067}{{\tt hep-th/0103067}}.

\bibitem{DijkgraafVafaI}
R.~Dijkgraaf and C.~Vafa, ``Matrix Models, Topological Strings, and
  Supersymmetric Gauge Theories,'' {\em Nucl. Phys. B} {\bf 644} (2002) 3--20,
  \href{http://xxx.lanl.gov/abs/hep-th/0206255}{{\tt hep-th/0206255}}.

\bibitem{DijkgraafVafaII}
R.~Dijkgraaf and C.~Vafa, ``On Geometry and Matrix Models,'' {\em Nucl. Phys.
  B} {\bf 644} (2002) 21--39,
  \href{http://xxx.lanl.gov/abs/hep-th/0207106}{{\tt hep-th/0207106}}.

\bibitem{DijkgraafVafaIII}
R.~Dijkgraaf and C.~Vafa, ``A Perturbative Window into Non-Perturbative
  Physics,'' \href{http://xxx.lanl.gov/abs/hep-th/0208048}{{\tt
  hep-th/0208048}}.

\bibitem{WittenLargeNChirDyn}
E.~Witten, ``Large N Chiral Dynamics,'' {\em Ann. Phys.} {\bf 128} (1980)
  363--375.

\bibitem{WittenthetaSolution}
E.~Witten, ``Theta Dependence in The Large $N$ Limit Of Four-Dimensional Gauge
  Theories,'' {\em Phys. Rev. Lett.} {\bf 81} (1998) 2862--2865,
  \href{http://xxx.lanl.gov/abs/hep-th/9807109}{{\tt hep-th/9807109}}.

\bibitem{ShifmanTunneling}
M.~A. Shifman, ``Domain walls and the decay rate of the excited vacua in large
  $N$ Yang-Mills theory,'' {\em Phys. Rev. D} {\bf 59} (1999) 021501,
  \href{http://xxx.lanl.gov/abs/hep-th/9809184}{{\tt hep-th/9809184}}.

\bibitem{GukovVafaWitten}
S.~Gukov, C.~Vafa, and E.~Witten, ``CFT's From Calabi-Yau Four-folds,'' {\em
  Nucl. Phys. B} {\bf 584} (2000) 69--108 [Erratum--ibid. B 608 (2001)
  477--478], \href{http://xxx.lanl.gov/abs/hep-th/9906070}{{\tt
  hep-th/9906070}}.

\bibitem{OoguriVafaWorldsheet}
H.~Ooguri and C.~Vafa, ``Worldsheet Derivation of a Large N Duality,'' {\em
  Nucl. Phys. B} {\bf 641} (2002) 3--34,
  \href{http://xxx.lanl.gov/abs/hep-th/0205297}{{\tt hep-th/0205297}}.

\bibitem{PerturbativeMatrixModels}
R.~Dijkgraaf, S.~Gukov, V.~A. Kazakov, and C.~Vafa, ``Perturbative Analysis of
  Gauged Matrix Models,'' {\em Phys. Rev. D} {\bf 68} (2003) 045007,
  \href{http://xxx.lanl.gov/abs/hep-th/0210238}{{\tt hep-th/0210238}}.

\bibitem{AttractorMech}
S.~Ferrara, R.~Kallosh, and A.~Strominger, ``$N=2$ Extremal Black Holes,'' {\em
  Phys. Rev. D} {\bf 52} (1995) 5412--5416,
  \href{http://xxx.lanl.gov/abs/hep-th/9508072}{{\tt hep-th/9508072}}.

\bibitem{AttStrom}
A.~Strominger, ``Macroscopic Entropy of $N=2$ Extremal Black Holes,'' {\em
  Phys. Lett. B} {\bf 383} (1996) 39--43,
  \href{http://xxx.lanl.gov/abs/hep-th/9602111}{{\tt hep-th/9602111}}.

\bibitem{AttFerrKallone}
S.~Ferrara and R.~Kallosh, ``Supersymmetry and Attractors,'' {\em Phys. Rev. D}
  {\bf 54} (1996) 1514--1524,
  \href{http://xxx.lanl.gov/abs/hep-th/9602136}{{\tt hep-th/9602136}}.

\bibitem{AttFerrKallTwo}
S.~Ferrara and R.~Kallosh, ``Universality of Supersymmetric Attractors,'' {\em
  Phys. Rev. D} {\bf 54} (1996) 1525--1534,
  \href{http://xxx.lanl.gov/abs/hep-th/9603090}{{\tt hep-th/9603090}}.

\bibitem{ColemanTunnel}
S.~R. Coleman, ``Fate Of the False Vacuum: Semiclassical Theory,'' {\em Phys.
  Rev. D} {\bf 15} (1977) 2929--2936 [Erratum--ibid. D. 16, 1248 (1977)].

\bibitem{WittenLargeNUONEproblem}
E.~Witten, ``Current Algebra Theorems For The $U(1)$ `Goldstone Boson','' {\em
  Nucl. Phys. B} {\bf 156} (1979) 269--283.

\bibitem{ShifmanthetaSolution}
M.~A. Shifman, ``Non-Perturbative Dynamics in Supersymmetric Gauge Theories,''
  {\em Prog. Part. Nucl. Phys.} {\bf 39} (1997) 1--116,
  \href{http://xxx.lanl.gov/abs/hep-th/9704114}{{\tt hep-th/9704114}}.

\bibitem{EvansthetaSolution}
N.~Evans, S.~D.~H. Hsu, and M.~Schwetz, ``Controlled Soft Breaking of $N=1$
  SQCD,'' {\em Phys. Lett. B} {\bf 404} (1997) 77--82,
  \href{http://xxx.lanl.gov/abs/hep-th/9703197}{{\tt hep-th/9703197}}.

\bibitem{HalpZhitTopSusc}
I.~Halperin and A.~Zhitnitsky, ``On Topological Susceptibility, Vacuum Energy
  and Theta,'' {\em Mod. Phys. Lett. A} {\bf 13} (1998) 1955--1967,
  \href{http://xxx.lanl.gov/abs/hep-ph/9707286}{{\tt hep-ph/9707286}}.

\bibitem{HalpZhitCanTheta}
I.~Halperin and A.~Zhitnitsky, ``Can $\theta/N$ Dependence for Gluodynamics be
  Compatible with $2\pi$ Periodicity in $\theta$?,'' {\em Phys. Rev. D} {\bf
  58} (1998) 054016, \href{http://xxx.lanl.gov/abs/hep-ph/9711398}{{\tt
  hep-ph/9711398}}.

\bibitem{HalpZhitIntegratingIn}
I.~Halperin and A.~Zhitnitsky, ````Integrating in'' and Effective Lagrangian
  for Non-Supersymmetric Yang-Mills Theory,'' {\em Nucl. Phys. B} {\bf 539}
  (1999) 166--186, \href{http://xxx.lanl.gov/abs/hep-th/9802095}{{\tt
  hep-th/9802095}}.

\bibitem{HalpZhitAnomalous}
I.~Halperin and A.~Zhitnitsky, ``Anomalous Effective Lagrangian and $\theta$
  Dependence in QCD at Finite $N_{c}$,'' {\em Phys. Rev. Lett.} {\bf 81} (1998)
  4071--4074, \href{http://xxx.lanl.gov/abs/hep-ph/9803301}{{\tt
  hep-ph/9803301}}.

\bibitem{ZhitDefects}
A.~Zhitnitsky, ``Topological defects and $\theta$ dependence in QCD,'' {\em
  Nucl. Phys. Proc. Suppl.} {\bf 73} (1999) 647--649.

\bibitem{StromingerConifold}
A.~Strominger, ``Massless Black Holes and Conifolds in String Theory,'' {\em
  Nucl. Phys. B} {\bf 451} (1995) 96--108,
  \href{http://xxx.lanl.gov/abs/hep-th/9504090}{{\tt hep-th/9504090}}.

\bibitem{GreeneMorrisonStromingerConifold}
B.~Greene, D.~Morrison, and A.~Strominger, ``Black Hole Condensation and the
  Unification of String Vacua,'' {\em Nucl. Phys. B} {\bf 451} (1995) 109--120,
  \href{http://xxx.lanl.gov/abs/hep-th/9504145}{{\tt hep-th/9504145}}.

\bibitem{HitchinGeneralized}
N.~Hitchin, ``Generalized Calabi-Yau Manifolds,'' {\em Q. J. Math.} {\bf 54}
  (2003) 281--308, \href{http://xxx.lanl.gov/abs/math.DG/0209099}{{\tt
  math.DG/0209099}}.

\bibitem{AshokDouglas}
S.~Ashok and M.~R. Douglas, ``Counting Flux Vacua,'' {\em JHEP} {\bf 0401}
  (2004) 060, \href{http://xxx.lanl.gov/abs/hep-th/0307049}{{\tt
  hep-th/0307049}}.

\end{thebibliography}\endgroup

\end{document}